\documentclass[12pt,preprint]{aastex}

\newcommand{\p}{\partial}
\newcommand{\e}{{\rm e}}
\newcommand{\dd}{{\rm d}}
\newcommand{\ie}{i.e. }
\newcommand{\eg}{e.g. }
\newcommand{\etal}{et al. }

\newcommand{\M}{{\cal M}}
\newcommand{\rsh}{{r}_{\rm sh}}
\newcommand{\sh}{{\rm sh}}
\newcommand{\Mc}{{{\cal M}^2}}
\newcommand{\hh}{h}

\shorttitle{advective-acoustic cycle}
\shortauthors{Foglizzo et al.}

\begin{document}

\title{Instability of a stalled accretion shock:  evidence for the advective-acoustic cycle }
\author{T. Foglizzo, P. Galletti}
\affil {Service d'Astrophysique, DSM/DAPNIA, UMR AIM CEA-CNRS-Univ. Paris VII,
CEA-Saclay, 91191 France}
\email{foglizzo@cea.fr}
\author{L. Scheck and H.-Th. Janka}
\affil{Max-Planck-Institut f\"ur Astrophysik, Karl-Schwarzschild-Str. 1, D-85741 Garching, Germany}
\author{ApJ received 2006 June 27; accepted 2006 September 20}

\begin{abstract}
We analyze the linear stability of a stalled accretion shock in a perfect gas with
a parametrized cooling function ${\cal L}\propto \rho^{\beta-\alpha} P^\alpha$.
The instability is dominated 
by the $l=1$ mode if the shock radius exceeds $2-3$ times the accretor
radius, depending on the parameters of the cooling function.
The growth rate and oscillation period are comparable to those observed 
in the numerical simulations of Blondin \& Mezzacappa (2006). \\
The instability mechanism is analyzed by separately measuring 
the efficiencies of the purely acoustic cycle 
and the advective-acoustic cycle. These efficiencies are estimated 
directly from the eigenspectrum, and also through a WKB analysis
in the high frequency limit. Both methods prove that the advective-acoustic 
cycle is unstable, and that the purely  acoustic cycle is stable. 
Extrapolating these results to low frequency leads us to interpret
the dominant mode as an advective-acoustic instability, 
different from the purely acoustic interpretation of Blondin \& Mezzacappa (2006).\\
A simplified characterization of the instability is proposed, based on an 
advective-acoustic cycle between the shock and the radius $r_\nabla$ 
where the velocity gradients of the stationary flow are strongest. 
The importance of the coupling region in this mechanism calls for a better 
understanding of the conditions for an efficient 
advective-acoustic coupling in a decelerated, nonadiabatic flow, 
in order to extend these results to core-collapse supernovae.

\end{abstract}


\section{Introduction}

The recent discovery of a strong $l=1$ instability of stalled accretion shocks in the context of core collapse supernovae (Blondin \etal 2003, Scheck \etal 2004, Ohnishi \etal 2006, Burrows \etal 2006) has revived the interest in the fundamental stability properties of accretion shocks. 
This instability could be a major ingredient in the mechanism of acceleration of neutron stars (Scheck \etal 2004, Janka \etal 2004, Scheck \etal 2006a).
It was also considered as a means to instigate g-mode dipole oscillations of the accreting neutron star (Burrows et al. 2006). While most of these authors recognized the presence of an advective-acoustic cycle similar to the one found by Foglizzo (2002, hereafter F02) in a different context, Blondin \& Mezzacappa (2006, hereafter BM06) challenged this interpretation and advocated a purely acoustic mechanism. Understanding the mechanism at work in this instability is a crucial step towards correctly extrapolating its consequences in a more realistic astrophysical situation.  

The physics of the advective-acoustic cycle is based on the linear coupling between acoustic and advected perturbations through the flow gradients: both entropy and vorticity perturbations act as source terms for the acoustic wave equation (Foglizzo 2001, hereafter F01, and Foglizzo \& Galletti 2003). 
This has been known for decades in the field of jet engines, since the pioneering works of Candel (1972), Howe (1975), Marble \& Candel (1977) and Abouseif et al. (1984). In the subsonic flow below the stalled shock, this linear coupling is due to the gradients associated with the convergence of the flow, its deceleration, gravity and cooling. The interaction between the shock and the flow gradients gives birth to an advective-acoustic cycle, in which an advected perturbation generates a pressure feedback which triggers, at the shock, a new advected perturbation. 
Although Foglizzo (2002) proved the instability of the advective-acoustic cycle in an accelerated isothermal flow, the fate of such cycles in a cooled decelerated flow is still an open question: can they account for the instability observed in the simulations of BM06 ?

The first aim of the present study is to clarify the instability mechanism at work, using perturbative techniques. The accretion flow is idealized as a perfect gas passing through a stationary shock, and subject to cooling processes schematically described by a cooling function ${\cal L}\propto \rho^{\beta-\alpha} P^\alpha$, mimicking in the simplest manner the neutrino cooling in the core-collapse context.
It is the first time that a linear approach has been used to 
understand the mechanism of this nonradial instability in the core-collapse context.

A first step is to confirm that the dominating $l=1$ mode identified by BM06 in the linear phase of their numerical simulation indeed corresponds to the most unstable eigenmode of the linear problem. 

Beyond the determination of the eigenspectrum and the validation of numerical simulations, we wish to 
address the question of the instability mechanism, using techniques similar to F02. These techniques allow for a direct interpretation of the full eigenspectrum in terms of the efficiencies of the acoustic cycle and the advective-acoustic cycle. Alternatively, these efficiencies can also be computed in the WKB approximation. We wish to use both methods in order to check whether the instability is of acoustic or advective-acoustic nature. Understanding the nature of the instability leads us to construct, in a companion paper (Foglizzo \etal 2006), a simple toy model which can be solved analytically and allows us to reach a fundamental understanding of some of the properties of the instability.

The present  paper is organized as follows: the boundary value problem associated with this stalled accretion shock is described in Sect.~\ref{sect_perturb}, where we establish the boundary conditions at the shock and compare them to those used by Houck \& Chevalier (1992, hereafter HC92) in the different context of supernova fallback. We determine in Sect.~\ref{sect_spectrum} the eigenfrequencies of the flows studied by BM06, and compare them with the linear phase of their numerical simulations. Then we investigate in Sect.~\ref{sect_mechan} the mechanism responsible for this instability using the techniques of F02. The purely acoustic cycle is shown to be stable, and the advective-acoustic cycle is shown to be unstable with respect to $l=1$ perturbations, in the range of validity of our approximations. These results are extrapolated to very low frequency perturbations in Sect.~\ref{sect_discuss}. The arguments of BM06 are reconciled with the advective-acoustic interpretation of the instability in Sect.~\ref{sect_BM06}. Results are summarized in Sect.~\ref{sect_conc}.

\section{Formulation of the eigenvalue problem\label{sect_perturb}}

\subsection{Description of the stationary flow\label{sect_stat}}
\begin{figure}
\plotone{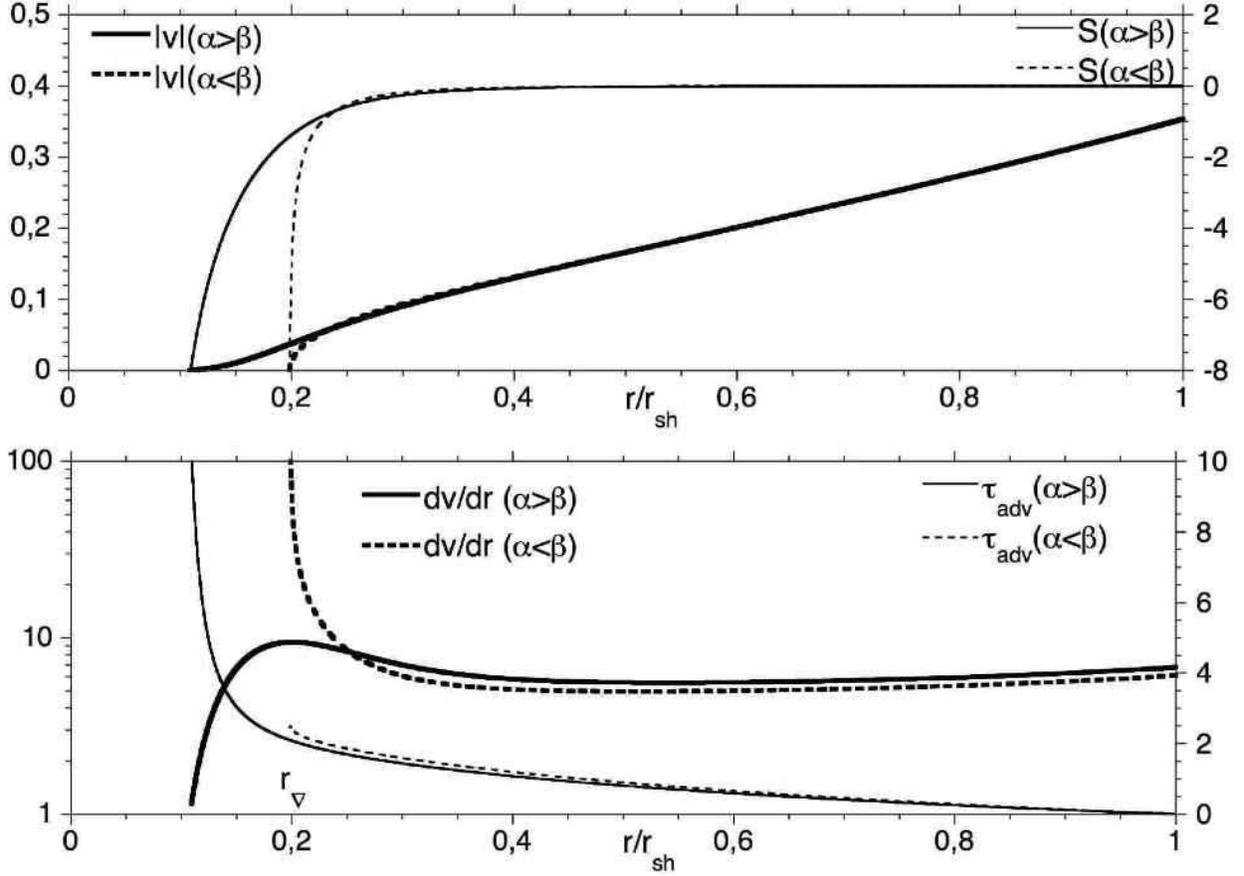}
\caption[]{Comparison of the stationary flows associated with the cooling parameters $\alpha=6$, $\beta=1$ (full lines) and $\alpha=3/2$, $\beta=5/2$ (dashed lines). The velocity and entropy profiles (upper plot) are similar in the outer parts of these two flows. The advection time $\tau_{\rm adv}$ from the shock to the accretor surface is finite if $\alpha<\beta$, and infinite if $\alpha>\beta$. The velocity gradient reaches a maximum at some intermediate height noted $r_\nabla>r_*$ if $\alpha>\beta$, whereas $r_\nabla=r_*$ if $\alpha<\beta$. The strength of cooling is chosen such that 
$r_\nabla /r_{\rm sh}=0.2$ in both flows.}
\label{figvS}
\end{figure}

We consider the radial accretion of a perfect gas with an adiabatic index $\gamma=4/3$, decelerated through a stationary shock at a radius $r_{\rm sh}$, accreting on the hard surface of a neutron star of mass $M$ and radius $r_*$. The self-gravity of the accreting gas is neglected. 
The cooling function ${\cal L}$ is defined as a parametrized function of density $\rho$ and 
pressure $P$ as in HC92:
\begin{eqnarray}
{\cal L}\propto \rho^{\beta-\alpha} P^\alpha\label{defL},
\end{eqnarray}
which allows us to mimic the effect of neutrino cooling using the same prescriptions $\alpha=3/2$, $\beta=5/2$ or $\alpha=6$, $\beta=1$ as BM06. Neutrino heating, and the associated effect of convection is ignored in the present study.\\
The equation of continuity, the Euler equation and the entropy equation defining the stationary flow between the shock and the accretor are written in spherical coordinates as follows:
\begin{eqnarray}
{\p \over\p r}(r^2\rho v) &=&0\ ,\label{eqcont}\\
{\p\over\p r}\left({v^2\over 2}+{c^2\over\gamma-1}-{GM\over r}\right)&=&
{{\cal L}\over \rho v}\ ,\label{eqbern}\\
{\p S\over \p r}&=&{{\cal L}\over Pv}\ ,\label{eqS}
\end{eqnarray}
where $S$ is a dimensionless measure of the entropy defined as the following function of pressure and density, normalized by their values $P_{\rm sh}$, $\rho_{\rm sh}$ immediately after the shock:
\begin{eqnarray}
S&\equiv&{1\over\gamma-1}\log\left\lbrack{P\over P_{\rm sh}}
\left({\rho_{\rm sh}\over\rho}\right)^\gamma\right\rbrack.\label{defS}
\end{eqnarray}
Note that the pressure force in the Euler equation (\ref{eqbern}) has been transformed using both this definition of $S$, and the sound speed $c$ defined by $c^2\equiv \gamma P/\rho$:
\begin{eqnarray}
{\nabla P\over\rho}=\nabla\left( {c^2\over\gamma-1}\right)-{c^2\over\gamma}\nabla S.\label{gradP}
\end{eqnarray}
The shock is assumed to be adiabatic. Following HC92 and BM06, we assume that the pre-shock velocity $v_1$ of the incoming gas is close to free fall: $v_1\sim v_{\rm ff}\equiv -(2GM/r_{\rm sh})^{1/2} $, and that the gas is cold: $\M_1\gg1$. The Mach number $\M\equiv |v|/c$ is defined as a positive number.

The assumption of stationarity required by the linear approach introduces a mathematical singularity at the surface $r_*$ of the accretor, where $v(r_*)=0$: the density diverges according to Eq.~(\ref{eqcont}) and the sound speed decreases to zero. Such pathologies are common in linear studies of cooled accretion on a hard surface (\eg from Chevalier \& Imamura 1982 to Saxton 2002), whatever the cooling function. For the cooling function considered here, two regimes can be distinguished depending on the sign of $\alpha-\beta$:
\par (i) If $\alpha-\beta<0$, the cooling efficiency increases as the gas cools down, leading to a cooling runaway. The potential energy is negligible compared to the cooling losses, and the advection time to the accretor surface is finite.
\par (ii) If $\alpha-\beta>0$, the cooling efficiency decreases as the gas cools down. The potential energy is comparable to the cooling losses, and the gas takes an infinite time to reach the surface.\\
These two regimes are illustrated in Fig.~\ref{figvS}, for the two set of cooling parameters used by 
BM06. The lower plot shows the velocity gradient of the stationary flow, which can participate to couple vorticity and acoustic perturbations (F02, Foglizzo \& Galletti 2003). The velocity gradient
reaches a maximum at a radius noted $r_\nabla$. Note that if $\alpha<\beta$, this maximum is reached on the accretor surface ($r_\nabla=r_*$). The strength of cooling in Fig.~\ref{figvS} was
chosen such that the $r_\sh/r_\nabla=5$ in both flows. As noticed by BM06, the two flows are very similar in their outer parts.\\
We find it convenient to use the variable $\log\M$ rather than the radius $r$ in order to solve numerically the differential system in the cooling layer near the accretor surface. Integration is stopped just above the accretor surface, when the Mach number reaches $10^{-9}$.

\subsection{Differential system ruling the perturbed flow}

The flow is perturbed in 3-D using spherical coordinates. The complex frequency $\omega\equiv (\omega_r,\omega_i)$ of the perturbations is defined such that its real part $\omega_r$ defines the oscillation frequency, and its imaginary part $\omega_i$ defines the growth rate.
The perturbation of velocity $\delta v_r,\delta v_\theta,\delta v_\varphi$, density $\delta \rho$, sound speed $\delta c$, and entropy $\delta S$ are used to define new perturbative functions $f$, $h$, $\delta K$, which enable a compact formulation of the differential system once projected on spherical harmonics $Y_l^m(\theta,\varphi)$:
\begin{eqnarray}
f&\equiv&{v\delta v_r}+{2\over\gamma-1}c\delta c,\label{deff}\\
g&\equiv&{\delta v_r\over v}+{2\over\gamma-1}{\delta c\over c},\label{defg}\\
\delta  K&\equiv& r^2v.\nabla\times\delta w+{l(l+1)c^2\over\gamma}
\delta S,\label{defKsph}
\end{eqnarray}
where $\delta w\equiv\nabla\times \delta v$ is the perturbation of vorticity. The resulting differential system is independent of the azimuthal number $m$:
\begin{eqnarray}
{\p f\over\p r}=\delta\left({{\cal L}\over \rho v}\right)\nonumber\\
+{i\omega v\over 1-\Mc}\left\lbrace
h -{f\over c^2} 
 +
\left\lbrack\gamma-1+{1\over\Mc}\right\rbrack{\delta S\over\gamma}
 \right\rbrace,\label{dfsp}
\\
{\p{ h}\over\p r}={i\delta K\over\omega r^2v}\nonumber\\
+{i\omega\over v(1-\Mc)}\left\lbrace
\frac{\mu^{2} }{c^{2}} f
 -\Mc { \hh}
- \delta S\right\rbrace,\label{dhsp}
\\
{\p \delta S\over\p r}={i\omega\over v}\delta S
+\delta\left({{\cal L}\over pv}\right),\label{dssp}
\\
{\p\delta K\over \p r}={i\omega\over v}\delta K
+l(l+1)\delta\left({{\cal L}\over \rho v}\right),\label{dksp}
\end{eqnarray}
where the quantity $\mu^2$ used in Eq.~(\ref{dhsp}) is defined by:
\begin{eqnarray}
\mu^2&\equiv&1-{l(l+1)c^2\over\omega^2r^2}(1-\Mc).\label{defmu}
\end{eqnarray}
The lengthy equations describing how the functions $f$, $h$, $\delta S$, $\delta K$ translate into the classical quantities $\delta v$, $\delta \rho$, $\delta P$, and the explicit expression of 
$\delta({\cal L}/\rho v)$ and $\delta({\cal L}/P v)$ in terms of $f,h,\delta S$, are written in Appendix~A.

\subsection{Boundary conditions at the shock}

The boundary conditions are established by writing the conservation laws in the frame of the perturbed shock. The derivation of these boundary conditions is shown in detail in Appendix~B, especially since we do not use the special system of spatial coordinates used by HC92. These moving spatial coordinates were introduced in perturbative studies of accretion shocks on white dwarfs (Chevalier \& Imamura 1982), such that the shock coordinate remains fixed 
even when the flow is perturbed. By contrast, our boundary conditions are expressed at the radius $r_\sh$ of the unperturbed shock, as functions of the displacement $\Delta\zeta$ of the shock, associated with its velocity $\Delta v\equiv -i\omega\Delta\zeta$. 
If the shock is strong ($\M_{\rm sh}^2=(\gamma-1)/2\gamma$) and the incoming gas is in free fall ($v_1=v_{\rm ff}$):
\begin{eqnarray}
{f_{\rm sh}\over c_{\rm sh}^2}&=&-{1\over\gamma}{\Delta v\over v_{\rm sh}} -\Delta \zeta
 {\nabla{S}_{\rm sh}\over \gamma} ,\label{fsh2}\\
h_{\rm sh}&=&{2\over\gamma+1}{\Delta v\over v_{\rm sh}} ,\label{hsh2}\\
\delta S_{\rm sh}&=&-\Delta \zeta
\left\lbrack \nabla S_\sh+
{1\over2r_{\rm sh}}
{5-3\gamma\over\gamma-1}
\right\rbrack-{2\over \gamma+1}{ \Delta v\over v_{\rm sh}}, \label{Ssh2}\\
\delta K_{\rm sh}&=&
-l(l+1)\Delta \zeta
\frac{c_{\rm sh}^{2}}{\gamma} \nabla{S}_{\rm sh} .\label{Kshsp}
\end{eqnarray}
These conditions can be translated into the classical quantities $\delta v_r$, $\delta \rho$, $\delta P$:
\begin{eqnarray}
\left({\delta v_r\over v}\right)_\sh={2\over\gamma+1}{\Delta v\over v_\sh}\nonumber\\
-{2\Delta\zeta\over\gamma+1}\left\lbrack
(\gamma-1)\nabla S_\sh+{\gamma\over2r_{\rm sh}}{5-3\gamma\over \gamma-1}\right\rbrack,\label{dvrsh}\\
\left({\delta \rho\over \rho}\right)_\sh={2\Delta\zeta\over\gamma+1}
\left\lbrack
(\gamma-1)\nabla S_\sh+{\gamma\over2r_{\rm sh}}{5-3\gamma\over \gamma-1}\right\rbrack,\label{drhosh}\\
\left({\delta P\over P}\right)_\sh=-2{\gamma-1\over\gamma+1}{\Delta v\over v_\sh}\nonumber\\
+{\Delta\zeta\over\gamma+1}\left\lbrack
(\gamma-1)^2\nabla S_\sh+{1+\gamma^2\over 2r_{\rm sh}}{5-3\gamma\over \gamma-1}
\right\rbrack,\label{dpsh}
\end{eqnarray}
The transverse components $\delta v_\theta$, $\delta v_\varphi$ of the velocity perturbation after the shock are related to the nonspherical deformation of the shock through:
\begin{eqnarray}
(\delta v_\theta)_\sh&=&{v_1-v_{\rm sh}\over r_{\rm sh}}{\p\Delta\zeta\over\p\theta},\label{dvtheta}\\
(\delta v_\varphi)_\sh&=&{v_1-v_{\rm sh}\over r_{\rm sh}\sin\theta}{\p\Delta\zeta\over\p\varphi}.\label{dvphi}
\end{eqnarray}
The divergence of the transverse velocity can be projected on the spherical harmonics $Y_l^m$ and satisfies the following boundary condition for a strong shock:
\begin{eqnarray}
{r_{\rm sh}\over\sin\theta}\left\lbrack{\p\over\p\theta}(\sin\theta\delta v_\theta)+{\p\over\p\varphi}\delta v_\varphi\right\rbrack=\nonumber\\
-{2l(l+1)\over\gamma-1}v_{\rm sh}\Delta\zeta.\label{dash}
\end{eqnarray}
Equations (\ref{dvrsh}), (\ref{drhosh}) and (\ref{dpsh}) are in perfect agreement with Eqs.~(50), (52) and (53) of HC92, which can be recovered by taking into account the gradients of the stationary flow quantities $v$, $\rho$ and $P$, using a Taylor expansion between $r_\sh$ and $r_\sh+\Delta\zeta$.\\
Transverse velocities at the shock are precluded by Eq.~(51) of HC92, in contradiction with our Eqs.~(\ref{dvtheta}-\ref{dvphi}) and (\ref{dash}), and also with other linear studies involving transverse perturbations such as Bertshinger (1986), Imamura \etal (1996), or Saxton \& Wu (1999).\\
As a consequence, the stability results reported by HC92 for nonradial perturbations should be considered as questionable, while their results concerning radial perturbations should be unaffected. 

\section{Numerical determination of the eigenfrequencies\label{sect_spectrum}}

\subsection{Numerical method\label{sect_prescription}}

\begin{figure}
\plotone{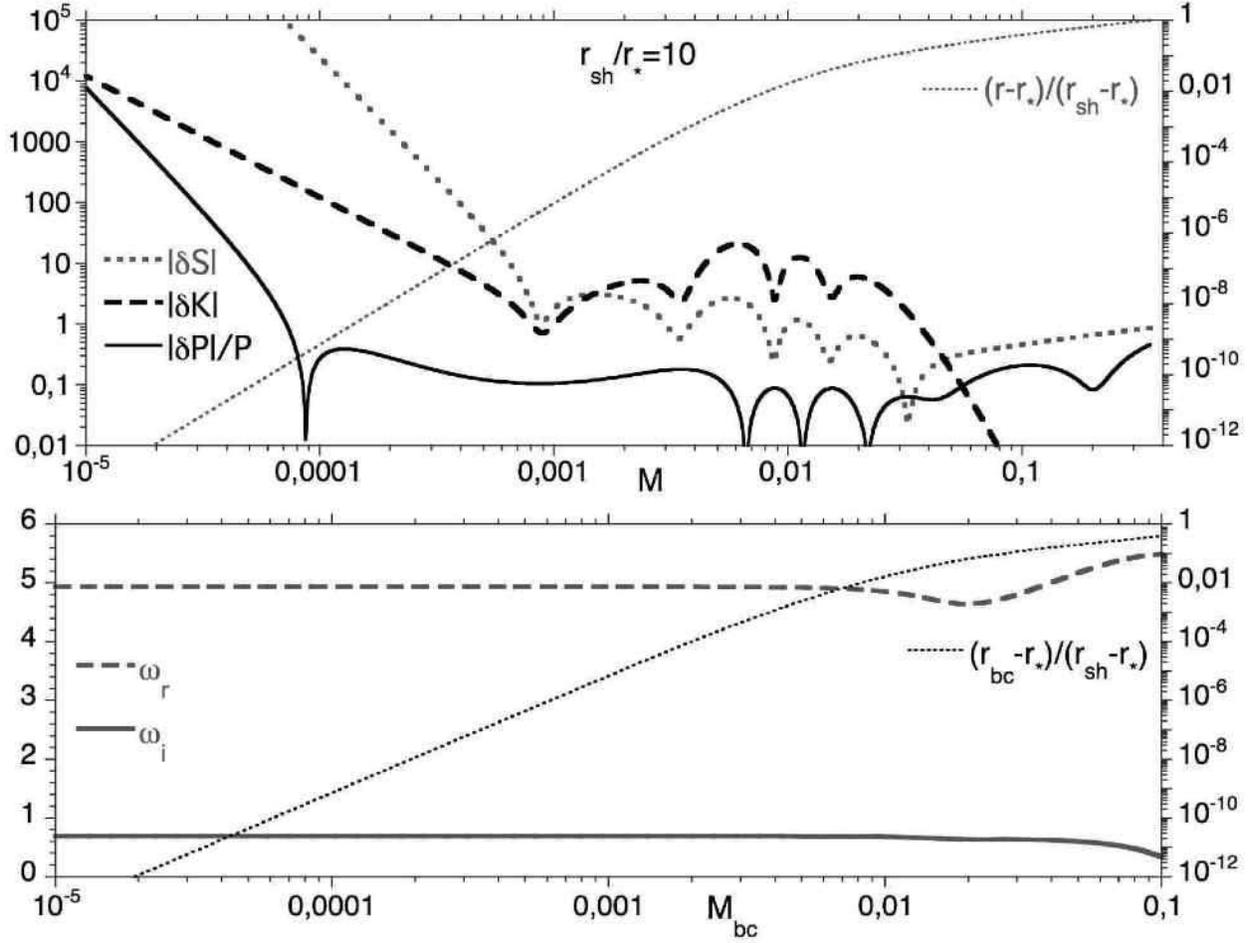}
\caption[]{In the upper plot, radial profiles of the perturbations of $|\delta K|$, $|\delta S|$ and $|\delta P|/P$ for the most unstable $l=1$ mode, in the flow with $\alpha=3/2$, $\beta=5/2$, $r_\sh/r_*=10$. The radial profiles are shown as a function of the Mach number $\M$, with the corresponding value of the
fractional radius shown on the right axis. The lower plot shows the influence of the radius $r_{\rm bc}$, and associated Mach number $\M_{\rm bc}$, of the lower boundary, on the complex eigenfrequency $\omega$: the very fine structure of the perturbations for $\M<4\times10^{-3}$ can be safely neglected. }
\label{fig_struct}
\end{figure}

The differential system is solved by integrating over the variable $\log\M$ from the shock down to the accretor surface, at a point where $\M=10^{-9}$ if the advection time is finite ($\alpha<\beta$). As an illustration, the radial shape of the eigenfunctions associated with the most unstable $l=1$ mode in the flow with $\alpha=3/2$, $\beta=5/2$, $r_\sh/r_*=10$, is shown in the upper plot of Fig.~\ref{fig_struct}. Using $\log\M$ as a variable allows us to compute the eigenfunctions down to the singular accretor surface. Structures are visible down to $\M\sim10^{-4}$, corresponding to radial scales which are much too small to be accessible to existing numerical simulations ($<10^{-9}r_\sh$). Luckily, these scales do not need to be resolved to measure the correct eigenfrequency, as shown in the lower plot of Fig.~\ref{fig_struct}. Varying the depth $r_{\rm bc}$ at which the boundary condition $\delta v(r_{\rm bc})=0$ is applied indicates that the regions of the flow where $\M<4\times 10^{-3}$ (\ie a fraction $<0.1\%$ of the shock distance) have a negligible effect on the eigenfrequency.\\
If the advection time is infinite ($\alpha\ge\beta$), the boundary condition $\delta v/v=0$ is applied at a radius where the advection time from the shock reaches 10 times the reference timescale $(r_\sh-r_*)/|v_\sh|$. We expect eigenmodes with a growth rate comparable to $|v_\sh|/(r_\sh-r_*)$ to be insensitive to any advective-acoustic artifact associated with this numerical prescription, because it occurs on a longer timescale than the instability. This expectation is validated by checking that the eigenfrequencies are unchanged by increasing the advection depth.\\
Once an eigenfrequency is found for a given intensity of the cooling function, it is tracked in the complex plane using the Newton-Raphson method. Eigenfrequencies are expressed in units of $|v_\sh|/(r_\sh-r_*)$ throughout the paper.

\subsection{Comparison with the eigenfrequencies estimated by BM06\label{sect_compare}}

\begin{figure}
\plotone{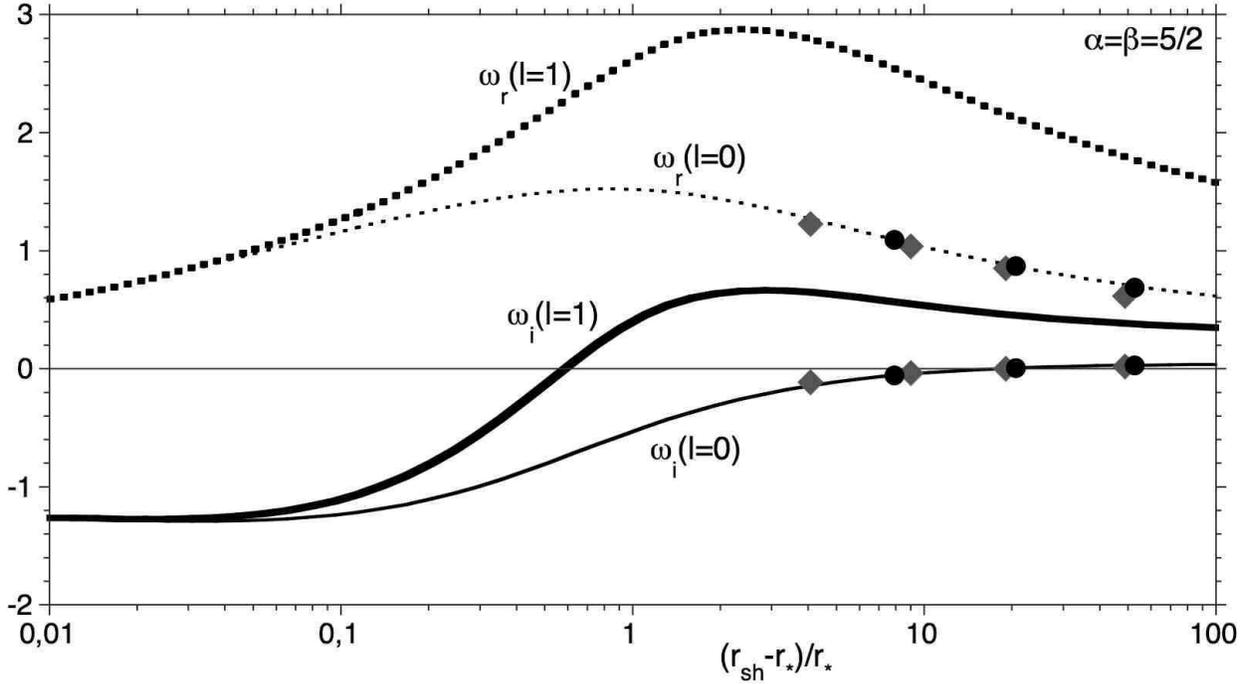}
\caption[]{Oscillation frequency $\omega_r$ and growth rate $\omega_i$ of the fundamental modes $l=0,1$, for $\alpha=\beta=5/2$. Also shown are the $l=0$ eigenfrequencies computed by HC92 (circles) and those measured in the 1-D simulation of BM06 (diamonds). The mode $l=1$ is always more unstable than the radial mode in this flow.}
\label{fig43a52b52}
\end{figure}

\begin{figure}
\plotone{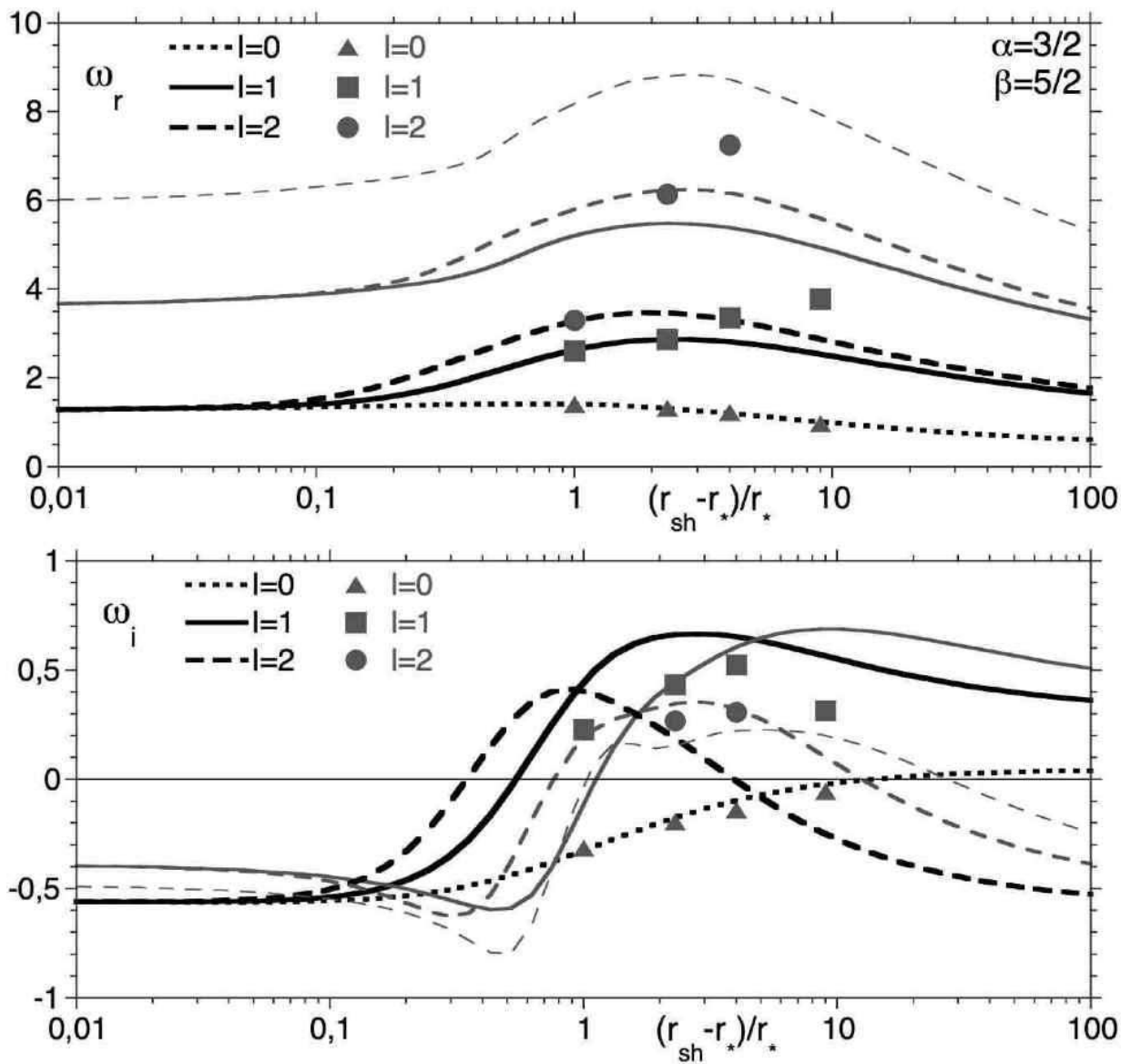}
\caption[]{Eigenfrequencies for a cooling law with $\alpha=3/2$ and $\beta=5/2$, corresponding to the modes $l=0$ (dotted line), $l=1$ (full line) and $l=2$ (dashed line). The fundamental mode is plotted with thick black lines. Harmonics are shown with thiner grey lines. The eigenfrequencies determined from BM06 are shown as triangles ($l=0$), squares ($l=1$) and circles ($l=2$).}
\label{fig43a32b52wri}
\end{figure}

\begin{figure}
\plotone{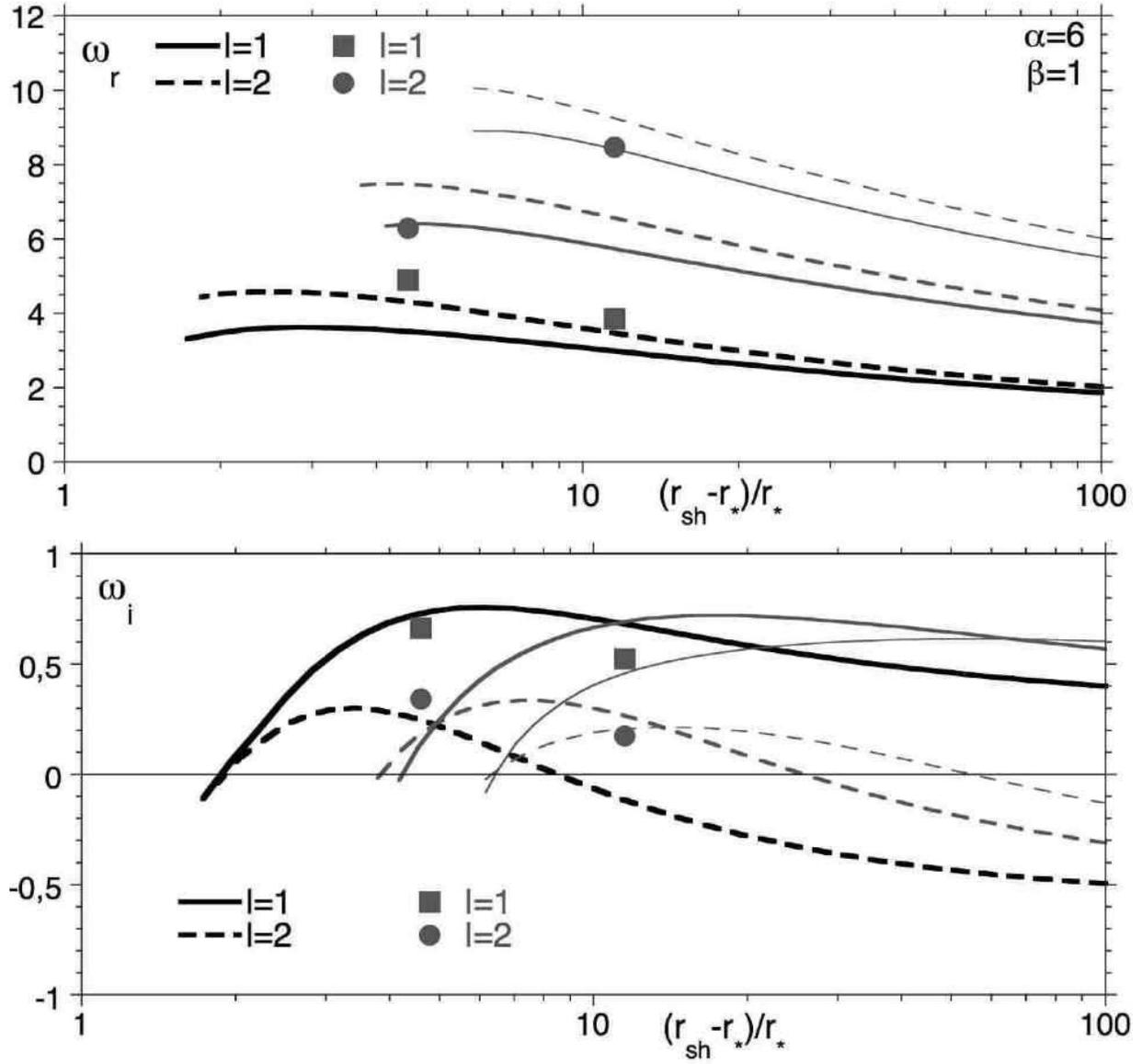}
\caption[]{Same as Fig.~\ref{fig43a32b52wri}, for the cooling law $\alpha=6$ and $\beta=1$.}
\label{fig43a6b1}
\end{figure}

\begin{figure}
\plotone{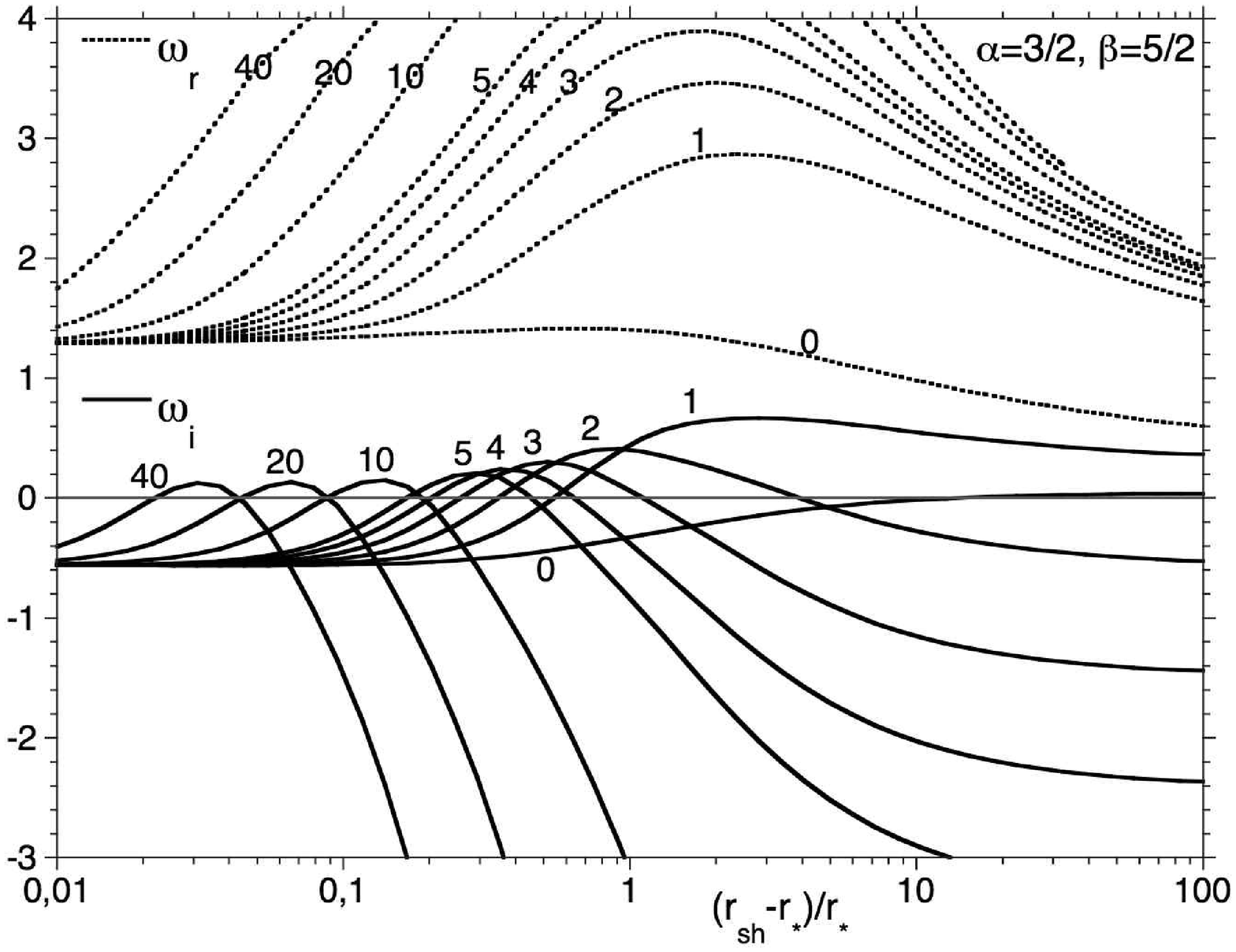}
\caption[]{Eigenfrequencies corresponding to the fundamental modes $0\le l\le 40$, for $\alpha=3/2$ and $\beta=5/2$. This plot indicates that the most unstable mode is always nonradial. The larger the shock radius, the smaller the degree $l\ge 1$ of the most unstable mode.}
\label{fig43a32b52l06}
\end{figure}

BM06 validated their numerical code in 1-D by comparing the eigenfrequency measured in the linear stage to the eigenfrequencies determined by HC92 for the mode $l=0$, with 
$\alpha=5/2$, $\beta=5/2$. Our calculation confirms this validation, as shown in 
Fig.~\ref{fig43a52b52}. Incidentally,  Fig.~\ref{fig43a52b52} shows that this flow is much more unstable to $l=1$ perturbations than to radial ones. \\
The nonspherical axisymmetric calculations of BM06 have been performed in a flow in which
$\alpha=3/2$, $\beta=5/2$, which was not considered by HC92. The eigenfrequencies of this flow are shown in Fig.~\ref{fig43a32b52wri}, with a globally acceptable agreement. 
The agreement seems significantly better for the $l=0$ mode than nonradial ones. 
Fig.~\ref{fig43a32b52wri} also shows that higher harmonics dominate the instability if the shock is far enough. For $r_\sh/r_*=5$, the instability of the mode $l=1$ should be dominated by the first harmonics, whereas the mode $l=2$ should be dominated by the second harmonics. \\
Fig.~\ref{fig43a6b1} shows the expected growth rate in the flow with $\alpha=6$, $\beta=1$ also considered by BM06, with a comparable agreement. Both Fig.~\ref{fig43a32b52wri} and 
\ref{fig43a6b1} enables us to evaluate the accuracy of numerical simulations: both the oscillation period and the growth time should be considered with a typical $30\%$ uncertainty.\\
Some of the discrepancies may be attributed to difficulties in disentangling higher harmonics which have a similar growth rate. Whether part of this discrepancy could be attributed to numerical viscosity or not depends to some extent on our understanding of the instability mechanism. At low frequency, a purely acoustic mechanism should be barely sensitive to numerical viscosity. By contrast a mechanism involving the advection of vorticity waves towards regions of small velocity may be more sensitive to numerical viscosity. This could contribute to explain why the growth rate measured in the nonradial numerical simulations seems systematically lower than that predicted by the linear analysis. 
The radial profiles of $\delta S$ and $\delta K$, displayed in Fig.~\ref{fig_struct}, show structures in the region of $\M\sim10^{-2}$ which cover a fraction $\sim2\%$ of the shock distance 
(see also the lower plot of Fig.~\ref{figpression}). Given the difficulty of advecting vorticity waves in a grid based code, a grid size of $0.1-0.2\%$ of the shock distance might be desirable near the accretor surface. Whether the 300-450 radial zones used in the simulations of BM06 are sufficient or not could be easily checked by performing new numerical simulations on a finer grid. 

BM06 have pointed out the resemblance between the instability in these two flows despite the different cooling function. Although we agree on the resemblance between these two flows when the shock radius is large, Fig.~\ref{fig43a6b1} reveals some significant differences for smaller shock radii: 
\par - the mode $l=1$ is always the most unstable mode if $\alpha=6$, $\beta=1$. By contrast, if $\alpha=3/2$, $\beta=5/2$, the instability may be dominated by perturbations with a higher degree $l\ge2$ if the shock radius is smaller than $2r_*$,
\par - the flow with $\alpha=6$, $\beta=1$ is stable if the shock radius is shorter than $\sim 2.5r_*$, whereas the flow with $\alpha=3/2$, $\beta=5/2$ is unstable whatever the shock distance.\\
A more detailed investigation of the instability at small shock distance when $\alpha=3/2$, $\beta=5/2$, illustrated by Fig.~\ref{fig43a32b52l06}, suggests that the most unstable mode corresponds to an azimuthal structure with a size comparable to the shock distance. \\
For both cooling functions, the dimensionless growth rate of the instability is at best of the order of $0.7-0.8$:
\begin{eqnarray}
\omega_i\le {|v_\sh|\over r_\sh-r_*}.
\end{eqnarray}
This similar growth rate can be viewed as a hint of a common physical mechanism of instability, which we investigate in the next section.

\section{Determination of the instability mechanism\label{sect_mechan}}

\subsection{Presence of oscillations in the eigenspectrum}

\begin{figure}
\plotone{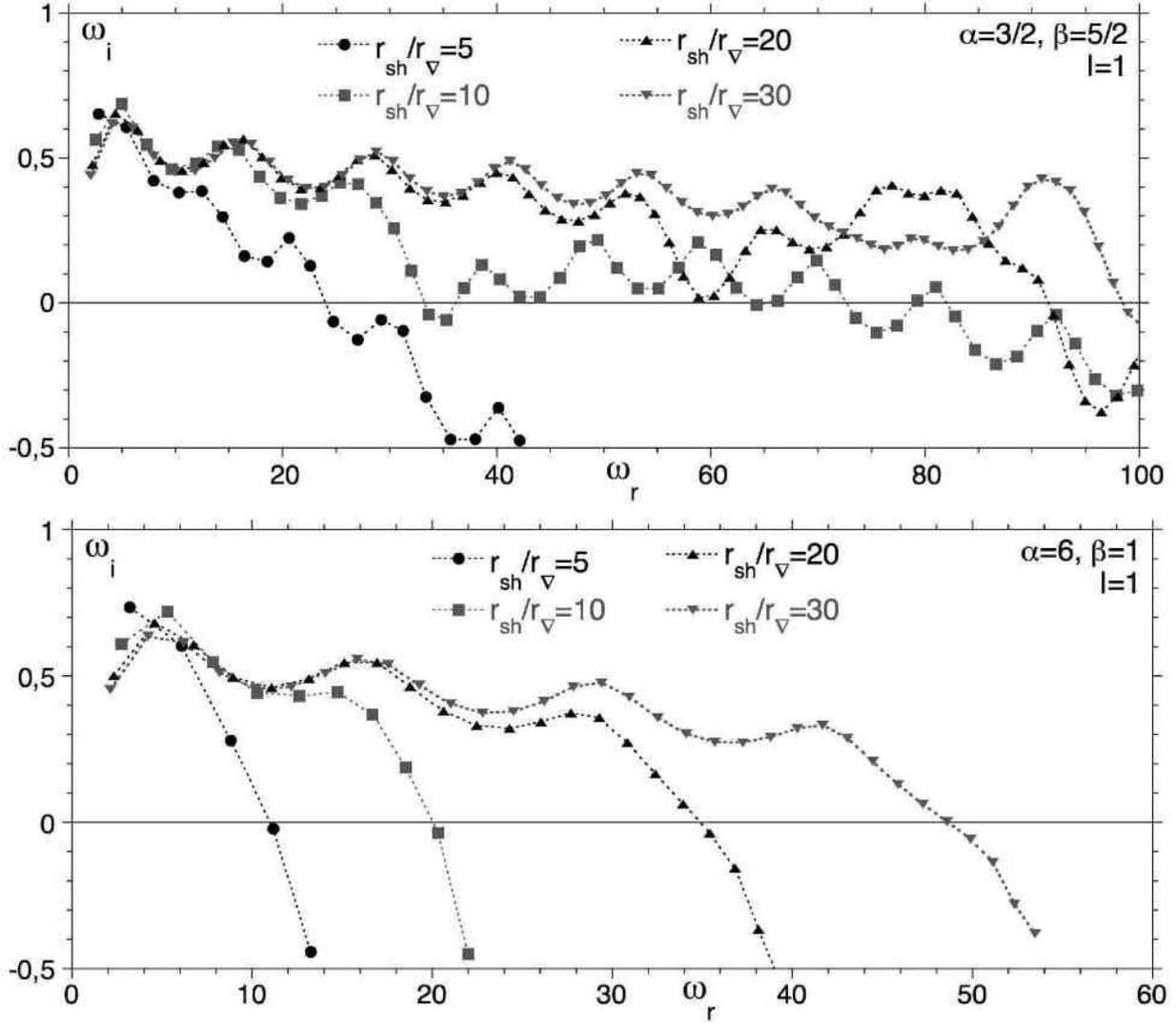}
\caption[]{Eigenfrequencies computed from the boundary value problem in a flow with $\alpha=3/2$, $\beta=5/2$ (upper plot) and $\alpha=6$, $\beta=1$ (lower plot), for different shock radii $r_\sh/r_\nabla$, for $l=1$ perturbations. }
\label{figr5_30}
\end{figure}

The oscillation period and growth time associated with the most unstable eigenmode, as determined in Figs.~\ref{fig43a32b52wri} and \ref{fig43a6b1}, should be a  signature of the instability mechanism. Unfortunately, our understanding of the possible instabilities is not deep enough to allow for a direct and conclusive interpretation of these timescales. The linear stability analysis can be helpful in determining the underlying mechanism, using the many other eigenfrequencies of the eigenspectrum. A global view of the eigenspectrum associated with $l=1$ perturbations is shown in Fig.~\ref{figr5_30} for both cooling functions, with different shock distances. Figure~\ref{figr5_30}  suggests that the larger the shock radius, the more numerous the unstable modes. A striking feature of these eigenspectra is the oscillations of the growth rate, which are easier to identify when the number of well defined eigenmodes is large, \ie when the shock radius is large. According to Fig.~\ref{figr5_30} , this identification requires $r_{\rm sh}/r_\nabla>10$ if $\alpha=6$, $\beta=1$, whereas the oscillations are already visible for $r_{\rm sh}/r_*=5$ if $\alpha=3/2$, $\beta=5/2$. These eigenspectrum oscillations are reminiscent of the oscillations visible in the eigenspectrum of the isothermal accretion flow accelerated towards a black hole (Fig.~4 of F02). They were explained by F02 as the consequence of the influence of a purely acoustic cycle interacting either constructively or destructively with the advective-acoustic cycle. The efficiencies of these two cycles are measured in the next Section, after recalling the formalism associated with these cycles.

\subsection{Calculation of the efficiencies ${\cal Q}$, ${\cal R}$ of the advective-acoustic and purely acoustic cycles\label{sect_formalism}}

\subsubsection{Advective-acoustic and purely acoustic cycles}

F02 developed a formalism in order to describe the advective-acoustic cycle, stable or not, below a stationary shock in a radial accretion flow onto a black-hole. 
The same formalism can be applied to a decelerated accretion flow onto a hard surface. Let us recall that vorticity and entropy perturbations are advected at the velocity of the flow, whereas pressure perturbations can propagate at the speed of sound. These two categories of perturbations would be linearly independent in a uniform flow, but are coupled linearly if the flow is inhomogeneous. The interaction between the stationary shock and the flow gradients gives birth to two cycles:

\par - an advective-acoustic cycle, whose duration is noted $\tau_{\cal Q}$. An advected perturbation of frequency $\omega_r$ generates a pressure feedback which triggers, at the shock, a new advected perturbation, whose amplitude has changed by a factor ${\cal Q}(\omega_r)$ after one advective-acoustic cycle. 
\par - a purely acoustic cycle , whose duration is noted $\tau_{\cal R}$. An acoustic perturbation of frequency $\omega_r$ propagating downward (not necessarily radially) produces a reflected (or refracted) perturbation reaching the shock and triggers a new pressure perturbation, whose amplitude has changed by a factor ${\cal R}(\omega_r)$ after one acoustic cycle.\\
The ``vortical-acoustic" instability studied by F02 is fundamentally nonradial because vorticity perturbations are the only advected perturbations in an isothermal flow. In a gas with $\gamma=4/3$, the
advective-acoustic cycles exist for both radial and non radial perturbations. A radial advective-acoustic cycle relies entirely on entropy perturbations. In an adiabatic flow, the acoustic feedback depends on the global increase of enthalpy in the postshock flow (Foglizzo \& Tagger 2000, F01) and can be efficient enough to destabilize the Bondi-Hoyle-Lyttleton accretion (Foglizzo, Galletti \& Ruffert 2005). The instability of this ``entropic-acoustic" cycle (\ie the mode $l=0$) is disfavoured in the core-collapse context because neutrino cooling precludes a strong adiabatic heating. 

\subsubsection{Eigenfrequencies associated with the cycles}

The simplest formulation of the advective-acoustic instability corresponds to a situation where the purely acoustic cycle is negligible. The instability threshold then corresponds to $|{\cal Q}|=1$ and the growth rate $\omega_i$ can be approximated by 
\begin{eqnarray}
\omega_i\sim{1\over\tau_{\cal Q}}\log|{\cal Q}|.\label{pureaa}
\end{eqnarray}
More generally, Foglizzo \& Tagger (2000) showed that the purely acoustic cycle is not necessarily negligible and modifies Eq.~(\ref{pureaa}) as follows:
\begin{eqnarray}
{\cal Q}\e^{i\omega\tau_{\cal Q}}+{\cal R}\e^{i\omega\tau_{\cal R}}=1.\label{dispQR}
\end{eqnarray}
This equation describing the simultaneous existence of two cycles $({\cal Q},\tau_{\cal Q})$ and
$({\cal R},\tau_{\cal R})$ is symmetric: it can account for an advective-acoustic instability ($|{\cal Q}|>1$) as well as an hypothetical acoustic instability (if $|{\cal R}|>1$). In the isothermal flow studied by F02, the acoustic cycle is ``weak" in the sense that the parameter $\epsilon<1$:
\begin{eqnarray}
\epsilon\equiv {|{\cal R}|\over |{\cal Q}|^{\tau_{\cal R}\over\tau_{\cal Q}}}<1,\label{smallR}
\end{eqnarray}
Assuming $\epsilon<1$, F02 showed from Eq.~(\ref{dispQR}) that the effect of the acoustic cycle is to either increase or decrease the growth raye $\omega_i$ in the following range:
\begin{eqnarray}
{1\over\tau_{\cal Q}}\log{|Q|\over 1+\epsilon}<\omega_i<{1\over\tau_{\cal Q}}\log{|Q|\over 1-\epsilon},\label{rangewi}
\end{eqnarray}
The case $\epsilon>1$ would be exactly symmetric, by exchanging $({\cal Q},\tau_{\cal Q})$ and $({\cal R},\tau_{\cal R})$ in Eqs.~(\ref{smallR}) and (\ref{rangewi}). 

\subsubsection{How to extract the cycles information directly from the eigenspectrum\label{sect_direct}}

\begin{figure}
\plotone{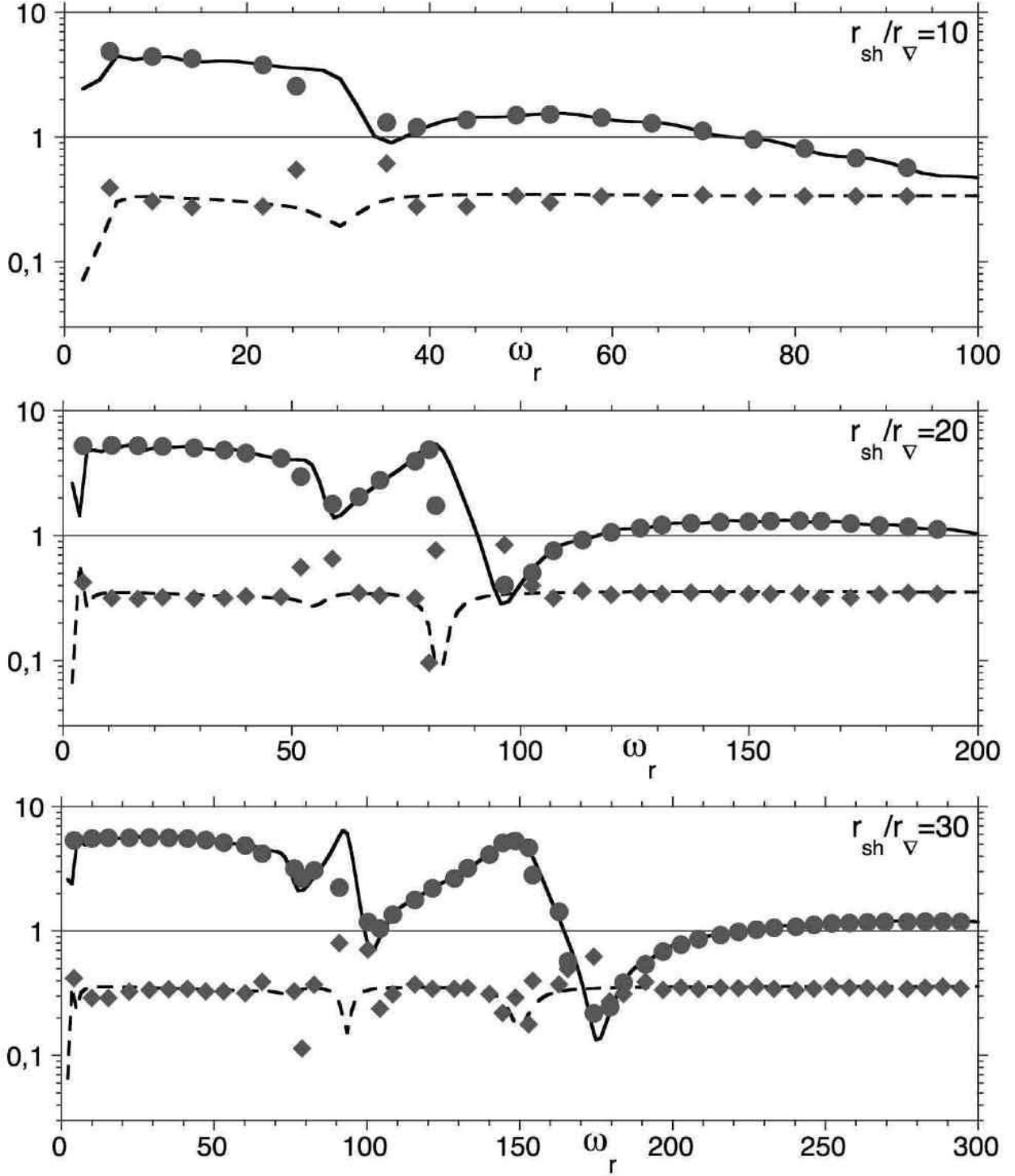}
\caption[]{Comparison between the coupling efficiencies $|{\cal Q}|(\omega)$, $|{\cal R}|(\omega)$ of the mode $l=1$, computed in the WKB approximation (full and dashed lines) and the ones deduced from the eigenspectrum (circles and diamonds) using Eqs.~(\ref{calQ}-\ref{calR}), in a flow with $\alpha=3/2$ and $\beta=5/2$, for different shock radii $r_\sh/r_*$. The agreement validates the formalism associated with the two cycles. The acoustic cycle is stable 
($|{\cal R}|<1$), and the advective-acoustic cycle can be unstable ($|{\cal Q}|\sim4-6$).}
\label{figa32b52QR}
\end{figure}

\begin{figure}
\plotone{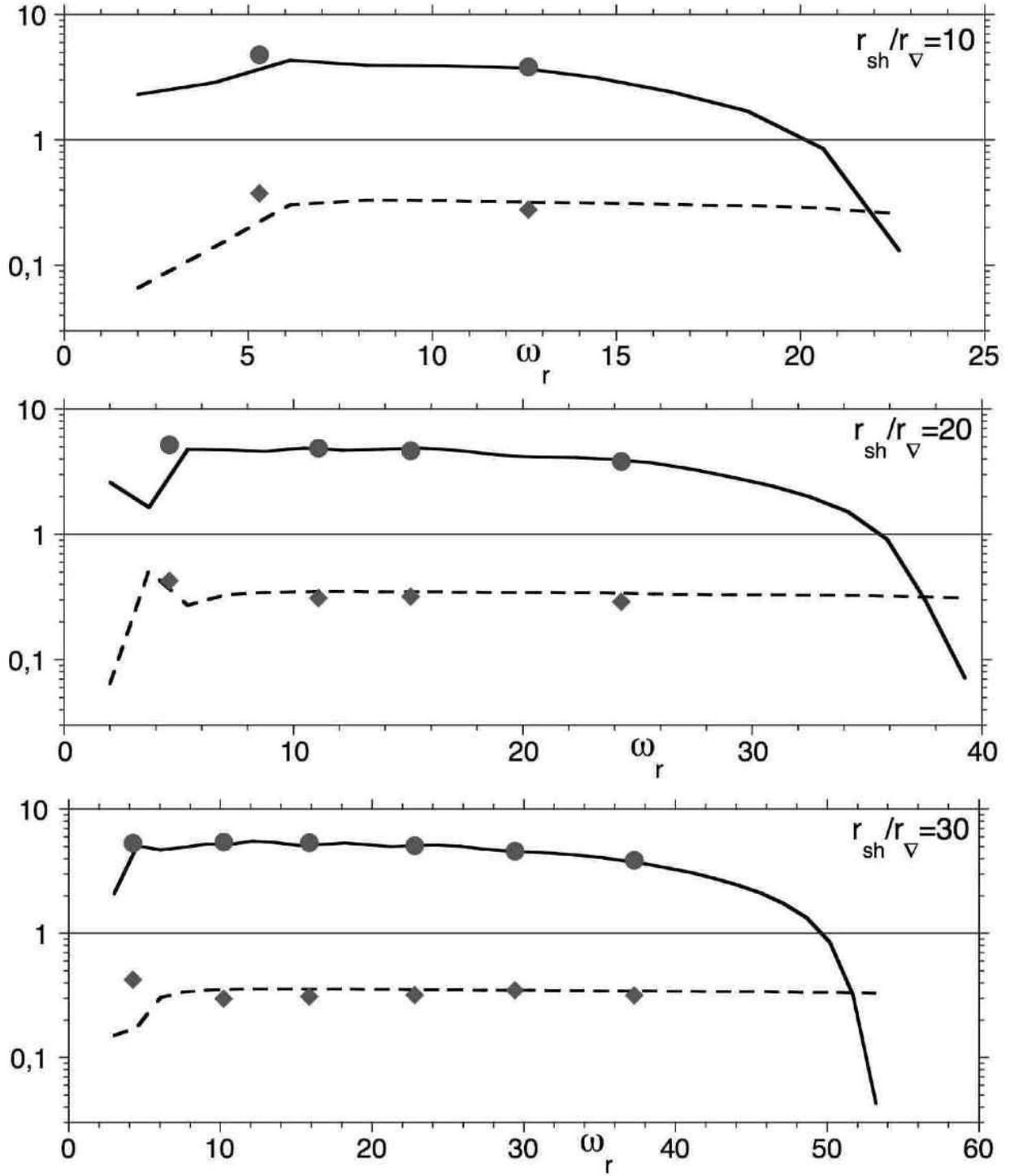}
\caption[]{Same as Fig.~\ref{figa32b52QR} with a different cooling function: $\alpha=6$ and $\beta=1$. The efficiencies $|{\cal Q}|$ and $|{\cal R}|$ are similar to those obtained with $\alpha=3/2$ and $\beta=5/2$. }
\label{figa6b1QR}
\end{figure}

Assuming that the underlying mechanism is due to a superposition of cycles described by 
Eq.~(\ref{dispQR}), an estimate of ${\cal Q}$, ${\cal R}$, $\tau_{\cal Q}$, $\tau_{\cal R}$ can be extracted directly from the oscillations observed in the eigenspectrum, in Fig.~\ref{figr5_30}. 
This method enables us to identify two cycles, one slow and one fast, and measure their efficiencies. For the sake of simplicity, we choose to denote the fast cycle with the letter ${\cal R}$, and the slow one with the letter ${\cal Q}$ (\ie $\tau_{\cal Q}>\tau_{\cal R}$).
The identification of the slow cycle with the advective-acoustic mechanism, and the fast cycle with the purely acoustic mechanism will become unambiguous in Sect.~\ref{sect_QRWKB}, which will also validate the assumption of an Eq.~(\ref{dispQR}) underlying the eigenspectrum.\\
According to F02, the timescale $\tau_{\cal Q}$ of the slowest cycle is related to the frequency difference $\omega_r({i+1})-\omega_r({i})$ between two consecutive eigenmodes, whereas the shortest timescale $\tau_{\cal R}$ is related to the frequency range $\Delta\omega_r$ of each oscillation. Denoting by $n_{\rm osc}$ the number of eigenmodes per oscillation,
\begin{eqnarray}
\tau_{\cal Q}&\sim&{2\pi\over\omega_r^{i+1}-\omega_r^{i}},\label{tqw}\\
{\tau_{\cal Q}\over \tau_{\cal R}}&\sim& n_{\rm osc}.\label{nosc}
\end{eqnarray}
For example, at first glance on the upper plot of 
Fig.~\ref{figr5_30}, one can anticipate that $\tau_{\cal Q}/\tau_{\cal R}\sim 5$  if $r_{\rm sh}/r_*=5$, whereas $\tau_{\cal Q}/\tau_{\cal R}\sim 10$ if $r_{\rm sh}/r_*=30$.\\
According to Eqs.~(54) and (55) of F02, the efficiencies $|{\cal Q}|$ and $|{\cal R}|$ associated with the slow and fast cycles respectively, can be extracted directly from the eigenspectrum, by measuring the 
following parameters of the eigenspectrum oscillations: (i) the frequency range $\Delta\omega_r$ of each oscillation, (ii) the number $n_{\rm osc}$ of eigenmodes per oscillation, (iii) the amplitude $\Delta\omega_{i}$ of the oscillations of the growth rate, and (iv) their average value ${\bar \omega}_{i}$:
\begin{eqnarray}
|{\cal Q}|&=&{\cosh \pi{\Delta\omega_{i}\over\Delta\omega_{r}}
\over
\cosh (n_{\rm osc}-1)\pi{\Delta\omega_{i}\over\Delta\omega_{r}}}
\exp2n_{\rm osc}\pi{{\bar\omega}_{i}\over\Delta\omega_{r}},\label{calQ}\\
|{\cal R}|&=&{\sinh n_{\rm osc}\pi{\Delta\omega_{i}\over\Delta\omega_{r}}
\over
\cosh (n_{\rm osc}-1)\pi{\Delta\omega_{i}\over\Delta\omega_{r}}}
\exp2\pi{{\bar\omega}_{i}\over\Delta\omega_{r}}.\label{calR}
\end{eqnarray}
The result of this method, applied to the $l=1$ eigenmodes of Fig.~\ref{figr5_30}, is shown in Fig.~\ref{figa32b52QR} for $\alpha<\beta$ and in Fig.~\ref{figa6b1QR} for $\alpha>\beta$. It shows that the fast cycle (diamonds) is always stable, whereas the slow cycle (circles) can be unstable. We now proceed to check the validity of these results by computing $|{\cal Q}|$ and $|{\cal R}|$ using another method, which will establish that the fast stable cycle is purely acoustic, and the slow cycle is advective-acoustic.

\subsubsection{An alternate way to measure $|{\cal Q}|$ and $|{\cal R}|$, in the WKB approximation\label{sect_QRWKB}}

Following F02, this method consists in determining the coupling efficiencies $|{\cal Q}|$ and $|{\cal R}|$ as continuous functions of the perturbation frequency $\omega_r$. The efficiency of the acoustic cycle is decomposed into ${\cal R}\equiv {\cal R}_{\rm sh}{\cal R}_{\nabla}$, while the efficiency of the advective-acoustic cycle is decomposed into ${\cal Q}\equiv {\cal Q}_{\rm sh}{\cal Q}_{\nabla}$, with ${\cal R}_{\rm sh}$, ${\cal R}_{\nabla}$, ${\cal Q}_{\rm sh}$ and ${\cal Q}_{\nabla}$ being defined as follows:
\par(i) when an acoustic wave of frequency $\omega_r$ propagating outward reaches the shock, ${\cal R}_{\rm sh}(\omega_r)$ measures the efficiency of acoustic reflection, while ${\cal Q}_{\rm sh}(\omega_r)$ measures the amount of advected perturbations (entropy/vorticity) produced by the shock.
\par (ii) when an acoustic wave of frequency $\omega_r$ propagates in the flow towards the accretor, ${\cal R}_{\nabla}(\omega_r)$ measures the amount of acoustic reflection. When an advected perturbation of frequency $\omega_r$ is advected towards the accretor, ${\cal Q}_{\nabla}(\omega_r)$ measures the amount of acoustic waves propagating against the flow. Note that ${\cal Q}_{\nabla}$ and ${\cal R}_{\nabla}$ are measured at a radius immediately below the shock, but do not involve the physics of the shock.\\
This approach is an extension of the approach used by F01, F02 in adiabatic and isothermal flows, where acoustic and advected perturbations are easily identified using a WKB approximation.
The technicalities of the method is described in Appendix~C and Appendix~D. The WKB approach assumes that the wavelength of the advected and acoustic perturbation is shorter than the scale of the flow gradients just below the shock. This method is thus expected to be reliable at high frequency, and to break down at low frequency.\\
Besides, the presence of cooling processes makes this method even more approximative, since we choose to neglect it in the immediate vicinity of the shock, for the sake of simplicity. Neglecting cooling is certainly justified when the shock is far enough from the accretor, as illustrated by the flat entropy profile in the upper plot of Fig.~\ref{figvS}, for $r_\sh/r_\nabla=5$. \\

The results of the WKB analysis applied to the $l=1$ mode are displayed as continuous and dashed lines in Figs.~\ref{figa32b52QR} and \ref{figa6b1QR}, together with the results of the first method. 
The agreement between the two methods is excellent at frequencies where $|{\cal Q}|$ and $|{\cal R}|$ vary smoothly with frequency. Since the method based on the eigenspectrum requires several neighboring eigenmodes to determine $|{\cal Q}|$ and $|{\cal R}|$, it is unable to correctly capture variations that are faster than the oscillation period $\Delta \omega_r$. This is particularly visible in the resulting estimate of $|{\cal R}|$, in Fig.~\ref{figa32b52QR}. \\
The direct method used in Sect.~\ref{sect_direct} can be used to estimate the range of validity of the WKB approximation. A close inspection of Figs.~\ref{figa32b52QR} and \ref{figa6b1QR} suggests that the WKB approximation is acceptable for $\omega_r\ge5$ in both flows. \\

Both methods indicate that the purely acoustic is stable, and is not even close to the instability threshold, with typical values $|{\cal R}|\le 0.5$. If a purely acoustic mechanism were responsible for the low frequency instability, as proposed by BM06, this mechanism would have to be stable at higher frequency
to be compatible with the results of our analysis.\\

More importantly, our calculations prove that the advective-acoustic cycle is unstable in the range of frequencies accessible to our analysis, with an efficiency reaching $|{\cal Q}|\sim 4-6$. The next sections aim at characterizing the properties of this instability, by estimating the timescale $\tau_{\rm Q}$ of the cycle and the effective coupling radius $r_{\rm eff}$ (Sect.~\ref{sect_tauq}), and finding approximations for its growth rate (Sect.~\ref{seuilwkb}) and oscillation period (Sect.~\ref{sect_oscil}). Whether this instability mechanism may be responsible for the dominant low frequency mode is discussed in Sect.~\ref{sect_discuss}.

\subsection{Estimate of the timescale $\tau_{\cal Q}$ and effective coupling radius $r_{\rm eff}$ of the advective-acoustic cycle\label{sect_tauq}}

\begin{figure}
\plotone{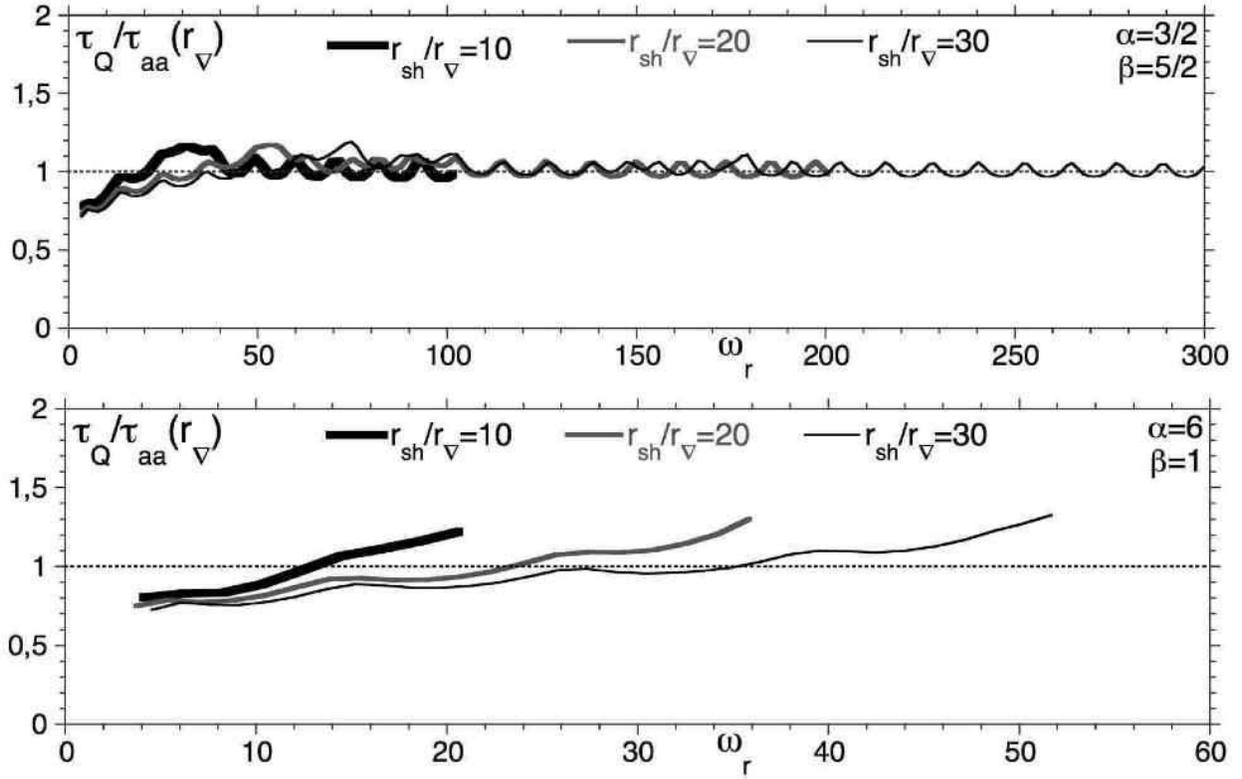}
\caption[]{Estimate of the cycle timescale $\tau_{\cal Q}$, directly extracted from the eigenspectrum 
(Eq.~\ref{tqw}). It is measured in units of the radial advective-acoustic time $\tau_\nabla\equiv\tau_{\rm aa}(r_\nabla)$, down to the radius $r_\nabla$ of maximum velocity gradient. At low frequency, the cycle timescale is 20\% shorter than $\tau_{\rm aa}(r_\nabla)$.}
\label{figtqtaa}
\end{figure}

\begin{figure}
\plotone{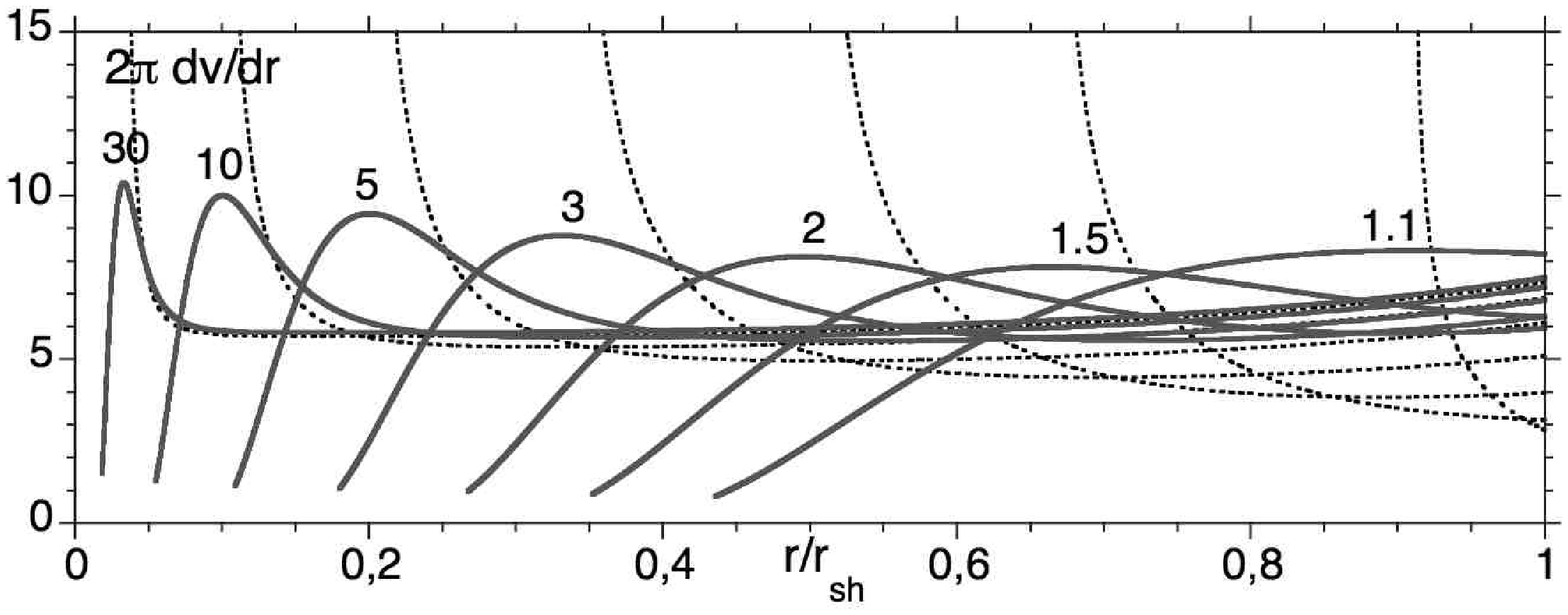}
\caption[]{Radial profile of the velocity gradient $2\pi \dd v/\dd r$, for different stationary flows, for both set of cooling parameters $\alpha=3/2$, $\beta=5/2$ (thin dotted lines) and $\alpha=6$, $\beta=1$ (full line). Each curve is labeled by the ratio $r_\sh/r_\nabla$. The velocity gradient is normalized  by $|v_\sh|/(r_\sh-r_*)$.}
\label{figdvdr}
\end{figure}

The accurate determination of the cycle timescales $\tau_{\cal Q}$ and $\tau_{\cal R}$, using the velocity and sound speed profiles of the stationary flow, is not straightforward: the duration $\tau_{\cal R}$ of the acoustic cycle depends on the depth of the turning point, and is thus a function of both the frequency $\omega_r$ and the order $l$. Similarly, $\omega_r$ and $l$ influence the depth at which an advected perturbation couples most efficiently to acoustic perturbations. The study of the advective-acoustic coupling in an adiabatic or isothermal flow showed that this coupling occurs all the way from the shock to the accretor (Eq.~(25) of F01, of Eq.~(29) of F02). Its net effect viewed from the shock radius may be summarized as a feedback from a single effective radius 
of coupling $r_{\rm eff}$, associated to an advective-acoustic timescale $\tau_{\cal Q}$. 
Let us define two radial reference timescales in the stationary flow:
\par (i) the radial acoustic timescale $\tau_{\rm ac}(r)$ from the shock to the radius $r$ and return:
\begin{eqnarray}
\tau_{\rm ac}(r)\equiv\int_{\rm sh}^r{2\over 1-\M^2}{\dd r\over c},\label{tac}
\end{eqnarray}
\par (ii) the radial advective-acoustic timescale $\tau_{\rm aa}(r)$ is defined as the advection timescale from the shock to the radius $r$, and acoustic return in the radial approximation:\\
\begin{eqnarray}
\tau_{\rm aa}(r)\equiv\int_{\rm sh}^r{1\over 1-\M}{\dd r\over \M c}.\label{taa}
\end{eqnarray}
The non radial character of the acoustic feedback has a minor influence on $\tau_{\rm aa}(r)$, given the subsonic character of the flow. By contrast, BM06 noted that the acoustic time for a very nonradial, purely acoustic wave can be significantly longer than along a radial path: the shock circumference is indeed a factor $\pi$ longer than its diameter.

In Fig.~\ref{figtqtaa}, we find it convenient to measure the cycle timescale $\tau_{\cal Q}$ deduced from Eq.~(\ref{tqw}), in units of $\tau_\nabla\equiv\tau_{\rm aa}(r_\nabla)$, where $r_\nabla$ is the radius defined in Sect.~\ref{sect_stat}, where the velocity gradient is maximum. The globally good matching between these two timescales indicates that velocity gradients are an important ingredient for the advective-acoustic coupling responsible for the acoustic feedback, as in the vortical-acoustic instability studied in a isothermal context by F02. Note that temperature gradients may also contribute to the advective-acoustic coupling, as seen in the adiabatic study of F01. 

We should keep in mind, however, that $r_\nabla$ is defined as a local maximum, whereas our definition of $r_{\rm eff}$ is global in the sense that it involves both the local coupling efficiency and the advection/propagation of the perturbations between the shock and $r_{\rm eff}$. In consequence, the approximation $r_{\rm eff}\sim r_\nabla$ should be viewed as a guide to our intuition rather than a fundamental property.

An advected perturbation of oscillation frequency $\omega_r$ is most sensitive to flow gradients whose lengthscale is shorter than the wavelength 
$2\pi v/\omega_r$. Those associated with velocity scale like $(\dd v/v\dd r)^{-1}$:
\begin{eqnarray}
\omega_r<2\pi {\dd v\over \dd r}.\label{seuilwdvdr}
\end{eqnarray}
According to Fig.~\ref{figdvdr}, the velocity gradient in the flow with $\alpha>\beta$ is very smooth and spread when the shock distance is short, whereas it gets sharper when the shock distance is large. This may explain, at least qualitatively,  why the flow with $\alpha=6$, $\beta=2$ is stable for $r_\sh/r_*<2.6$ whereas the flow with $\alpha=3/2$, $\beta=5/2$ is unstable even for a small shock distance. \\
Similarly, the divergence of the velocity gradient in the flow $\alpha<\beta$ is likely to couple advected and acoustic perturbations at much higher frequencies than in the flow $\alpha>\beta$, where the velocity gradient is smoother. This qualitative argument may explain why the 
range of unstable frequencies is so much larger for $\alpha=3/2$, $\beta=5/2$ than for $\alpha=6$, $\beta=1$ (as visible in Fig.~\ref{figr5_30} and summarized in Fig.~\ref{figa6b1wcut_wopt}).\\ 
Fig.~\ref{figdvdr} also illustrates the fact that velocity gradients are present all the way from the shock to $r_\nabla$, and may affect perturbations with a low frequency $\omega_r<5$ according to 
Eq.~(\ref{seuilwdvdr}). This threshold is comparable to our estimate of the threshold  of the WKB approximation in Sect.~\ref{sect_QRWKB}. \\
We interpret the $20\%$ offset visible in Fig.~\ref{figtqtaa} at low frequency as a consequence of the coupling through the various flow gradients, including velocity, above $r_\nabla$. Keeping in mind this low frequency distortion, the timescale $\tau_\nabla$ can be considered an acceptable approximation of $\tau_{\cal Q}$ for unstable $l=1$ modes.

\subsection{First order approximation of the growth rate\label{seuilwkb}}

\begin{figure}
\plotone{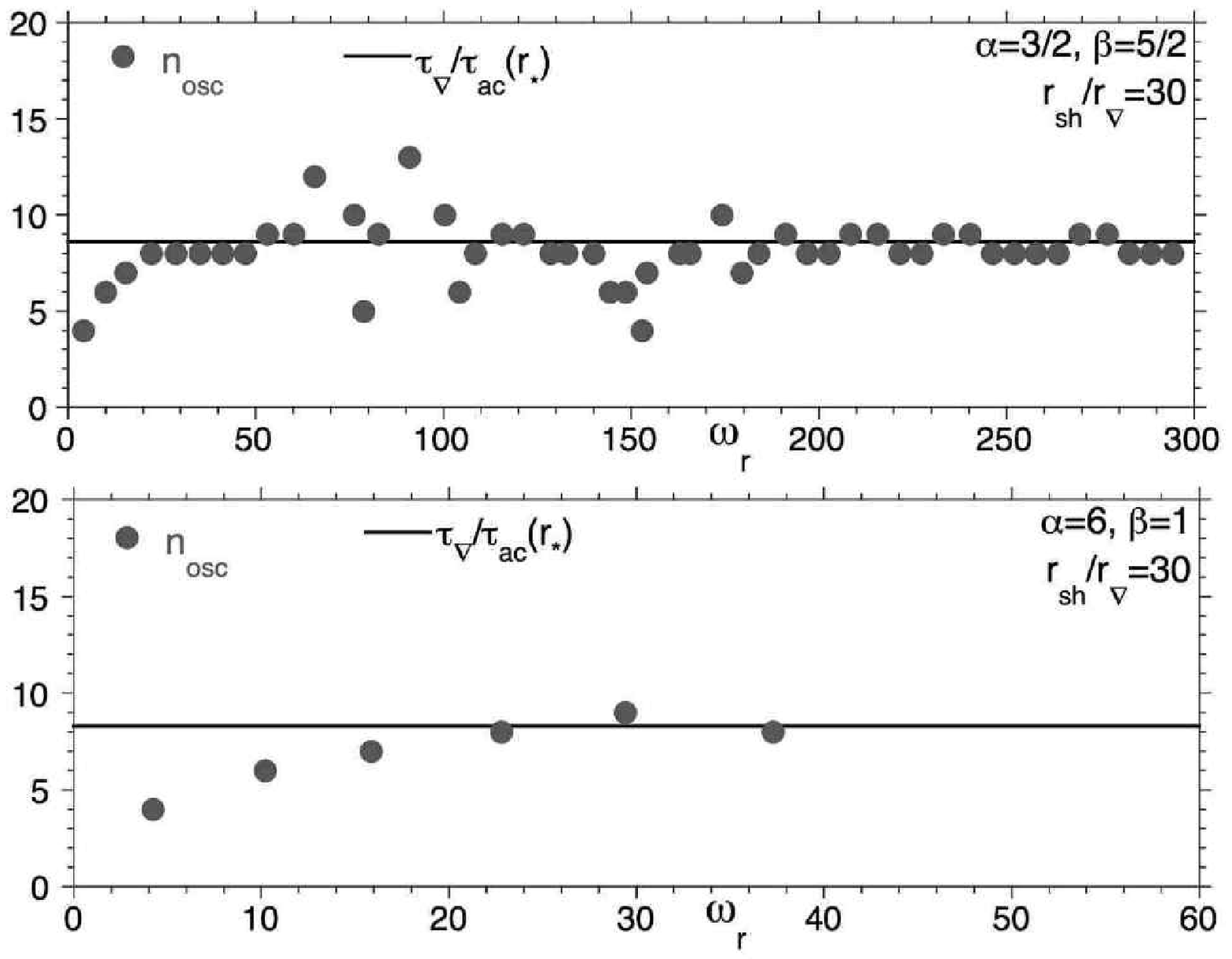}
\caption[]{Comparison between the number $n_{\rm osc}$ of eigenmodes per eigenspectrum oscillation, and the ratio of timescales $\tau_\nabla/\tau_{\rm ac}(r_*)$, for the two cooling functions, with $r_\sh/r_\nabla=30$. }
\label{figa32b52r29Nosc}
\end{figure}
\begin{figure}
\plotone{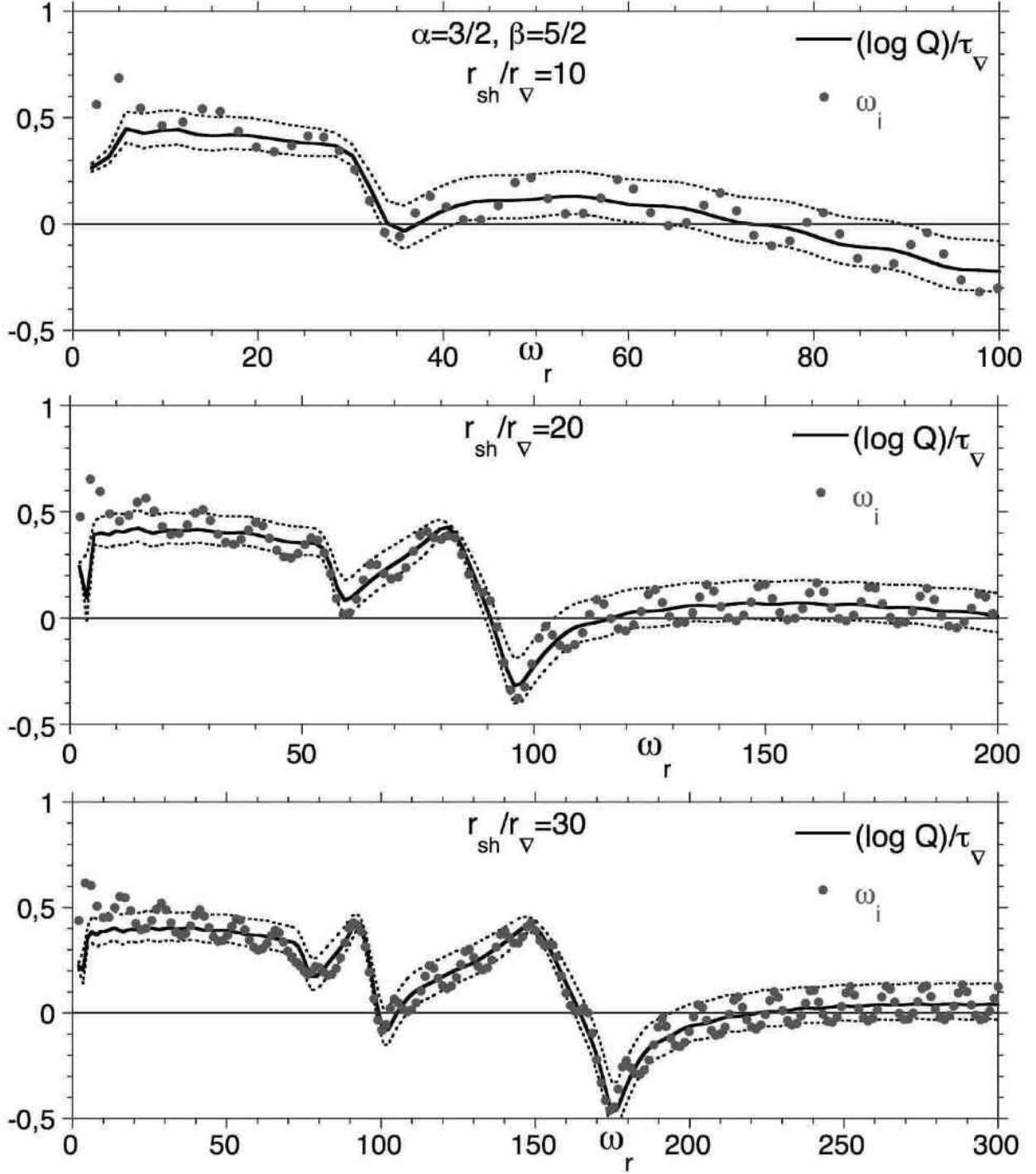}
\caption[]{Comparison between the eigenfrequencies computed from the boundary value problem and the first order estimate $(\log{\cal Q})/\tau_\nabla$, in a flow with $\alpha=3/2$ and $\beta=5/2$, for different shock distances $r_\sh/r_\nabla$. The dotted lines correspond to the minimum and maximum growth rates described by Eq.~(\ref{rangewi}), in which $\tau_{\cal Q}/\tau_{\cal R}$ is approximated by 
$\tau_\nabla/\tau_{\rm ac}(r_*)$.}
\label{figa32b52r9_29wi}
\end{figure}
\begin{figure}
\plotone{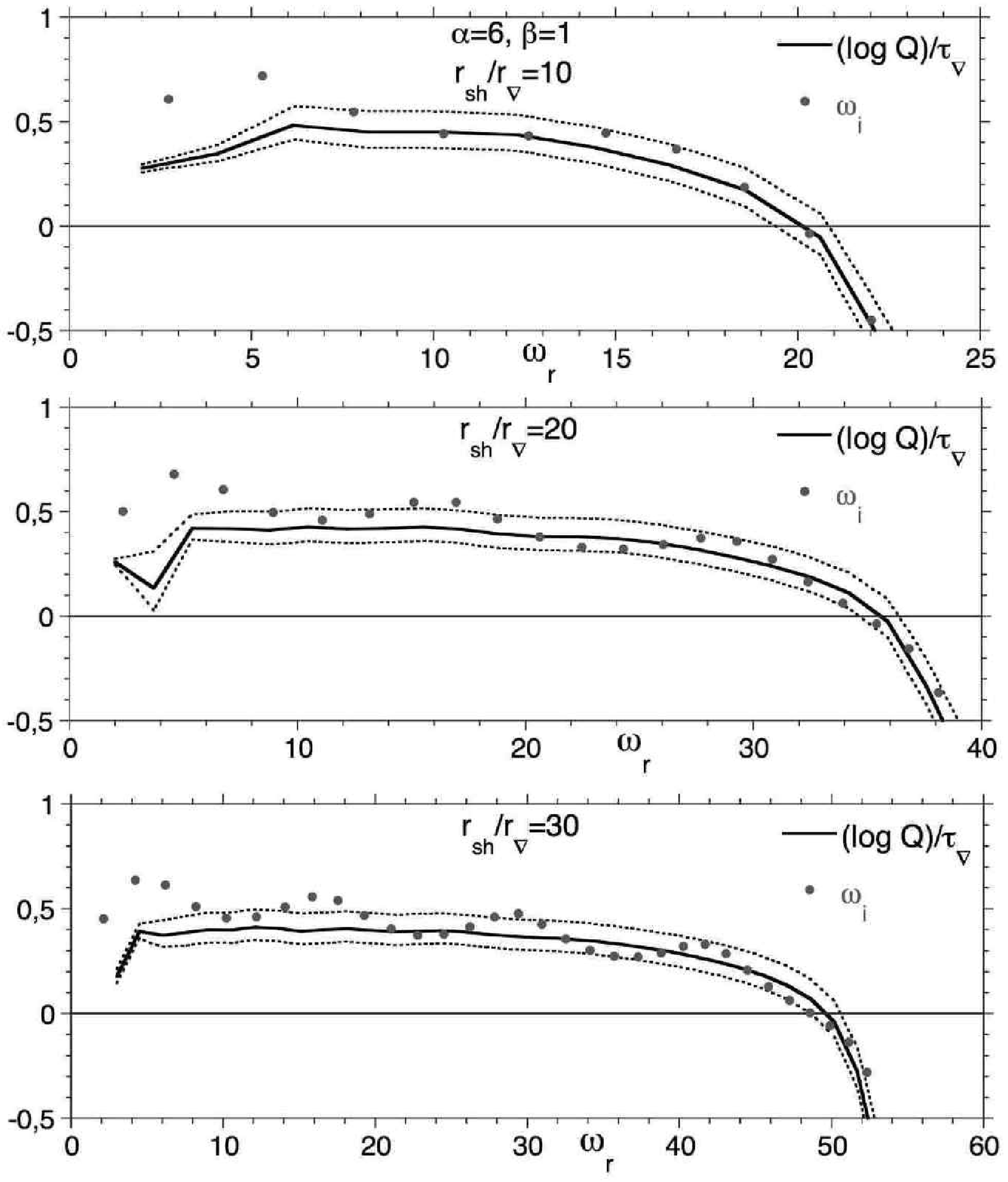}
\caption[]{Same as Fig.~\ref{figa32b52r9_29wi}, with a different cooling function: $\alpha=6$ and $\beta=1$. }
\label{figa6b1r10_30wi}
\end{figure}
Finding a discrete set of eigenmodes is a laborious task, in comparison with the straightforward calculation of $|{\cal Q}|$ and $|{\cal R}|$ in the WKB approximation. In this respect, Eq.~(\ref{rangewi}) provides us with a useful estimate of the growth rate if the parameters of the cycle $\tau_{\cal Q}$, $\tau_{\cal R}$, $|{\cal Q}|$ and $|{\cal R}|$ are known. Although our estimate $\tau_\nabla$ of $\tau_{\cal Q}$ is not fully satisfactory, we may continue along this direction in order to check what kind of accuracy can be reached. The effect of the acoustic cycle described by Eq.~(\ref{rangewi}) depends through $\epsilon$ on the ratio $\tau_{\cal Q}/\tau_{\cal R}$ (Eq.~\ref{smallR}), which is measured by $n_{\rm osc}$ 
(Eq.~\ref{nosc}). Despite the potentially significant difference between the timescales of radial and non radial acoustic modes, we choose to approximate $\tau_{\cal R}$ by $\tau_{\rm ac}(r_*)$, and compare
in Fig.~\ref{figa32b52r29Nosc} the number $n_{\rm osc}$ measured on the eigenspectra of 
Fig.~\ref{figr5_30} to the reference ratio $\tau_\nabla/\tau_{\rm ac}(r_*)$. The global agreement is acceptable given the discreteness of the number of nodes, the gross approximation of $\tau_{\cal R}$ by $\tau_{\rm ac}(r_*)$, and the difficulty of identifying the oscillation, especially when ${\cal Q}$ varies rapidly near 
$\omega_r\sim90$ and $\omega_r\sim 150$ for $\alpha=3/2$, $\beta=5/2$. The systematic  decrease of $n_{\rm osc}$ observed at low frequency is compatible with the decrease of $\tau_{\cal Q}/\tau_\nabla$ discussed in Sect.~\ref{sect_tauq}. \\

The growth rate $\omega_i$ measured in the eigenspectra of Fig.~\ref{figr5_30} is then compared in 
Figs.~\ref{figa32b52r9_29wi} and \ref{figa6b1r10_30wi} to the value $(\log{\cal Q})/\tau_\nabla$ (full line), and the expected range of influence of the acoustic cycle deduced from Eq.~(\ref{rangewi}) (dotted lines). This comparison is interpreted as follows:
\par (i) the amplitude of the eigenspectrum oscillation is very well matched by Eq.~(\ref{rangewi}),
\par(ii) the global shape of the eigenspectrum is globally very well reproduced
\par(iii) as expected, some systematic discrepancies are observed at low frequency, concerning the first $\sim10$ eigenmodes. This discrepancy can reach a factor 2 for the fundamental mode.\\
We first conclude that the radius $r_\nabla$ is an excellent approximation of the feedback radius at high frequency, for both cooling functions. This provides us with a rather simple description of the instability mechanism at work at high frequency.\\
The discrepancy observed at very low frequency exceeds the $20\%$ effect due to overestimating 
$\tau_{\cal Q}$ because the WKB approximation used to compute $|{\cal Q}|$ and $|{\cal R}|$ ceases to be valid, as already pointed out in Sect.~\ref{sect_QRWKB}.
Figs.~\ref{figa32b52r9_29wi} and \ref{figa6b1r10_30wi} suggest that the WKB approximation may only provide a gross estimate of the growth rate of the lowest frequency modes, within a factor 2.

\subsection{Oscillation timescale and efficiency $|{\cal Q}|$ associated with the most unstable mode\label{sect_oscil}}

\begin{figure}
\plotone{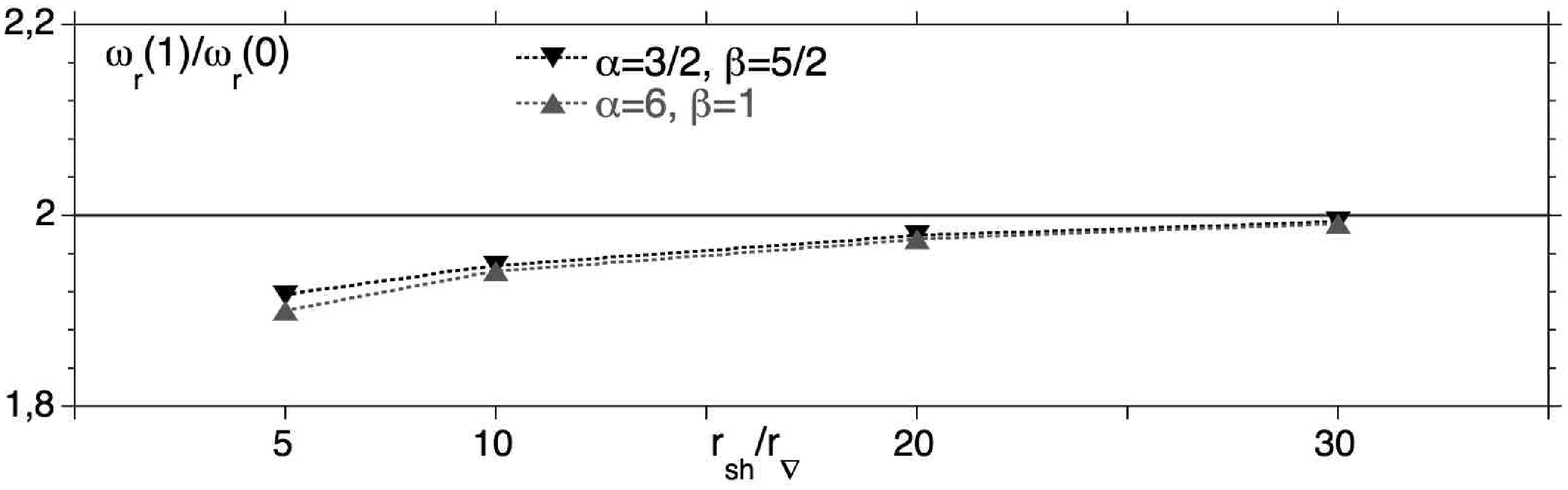}
\caption[]{Ratio of the frequencies of the first two eigenmodes $\omega_r(1)/\omega_r(0)$ corresponding to $l=1$ perturbations: the oscillation period $2\pi/\omega_r$ of the fundamental mode is an excellent measure of the advective-acoustic timescale $\tau_{\cal Q}$. }
\label{figphase}
\end{figure}
\begin{figure}
\plotone{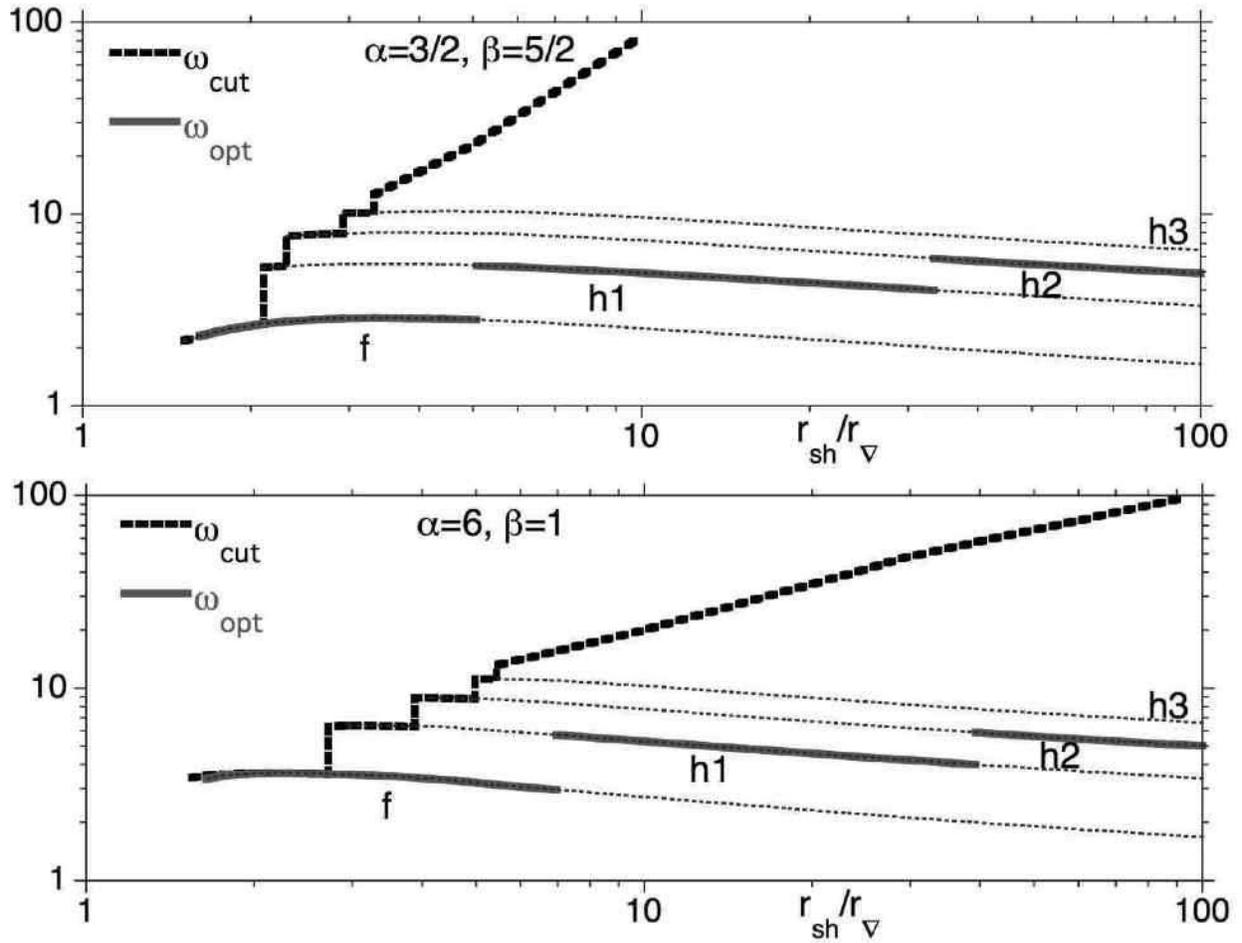}
\caption[]{Range of unstable frequencies of the $l=1$ mode. The frequency of the most unstable mode (thick full line) corresponds to one of the first harmonics noted ``f", ``h1", ``h2". The cut-off frequency (short dashed thick line) is a steeper function of the shock radius in the flow with $\alpha<\beta$. }
\label{figa6b1wcut_wopt}
\end{figure}
\begin{figure}
\plotone{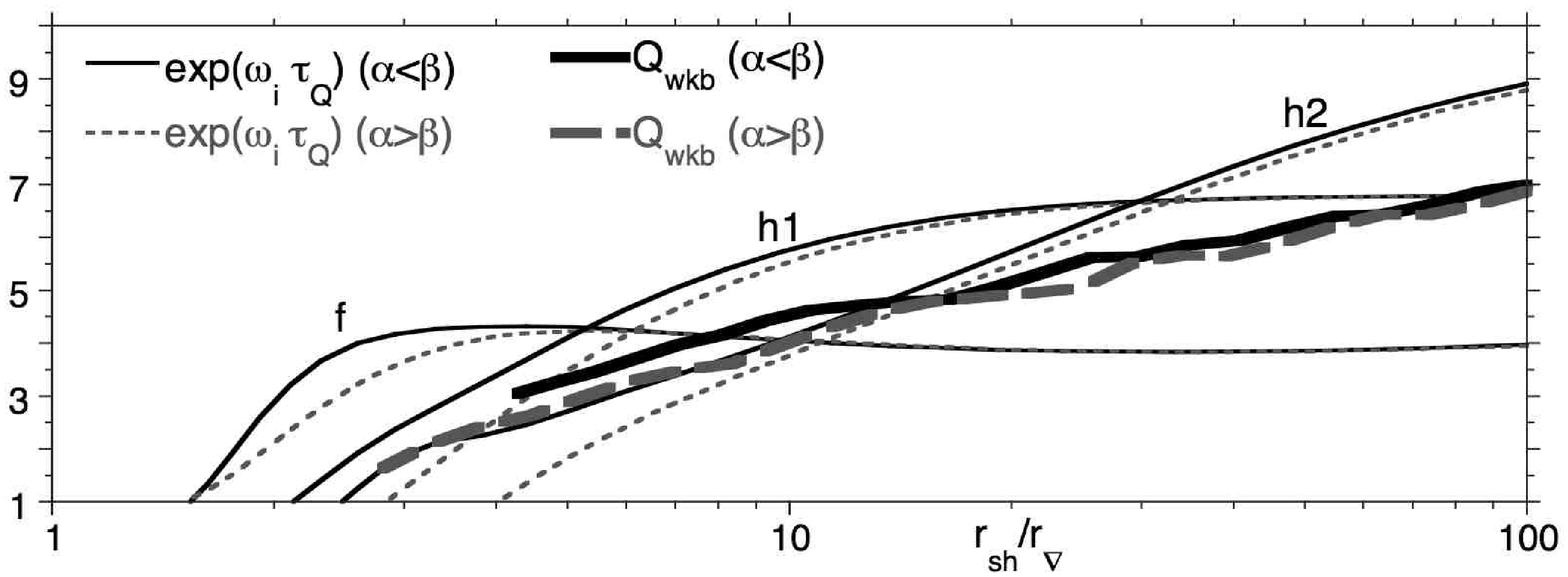}
\caption[]{Maximum efficiency $|{\cal Q}|_{\rm wkb}$ of $l=1$ perturbations in the WKB approximation (thick lines), compared to $\exp(\omega_i\tau_{\cal Q})$ of the low frequency eigenmodes ``f", ``h1", and ``h2". The cycle timescale is approximated as $\tau_{\cal Q}\sim 2\pi(k+1)/\omega_r(k)$. }
\label{figQ0}
\end{figure}

The relationship between the oscillation period of the fundamental mode and the timescale of the cycle is not obvious a priori, even if the acoustic cycle is neglected, because it depends on the phase $\phi_{\cal Q}$ of the complex efficiency ${\cal Q}$. Denoting by $\omega_r(k)$ the frequency of the $k$-th harmonic, the phase relation associated with Eq.~(\ref{dispQR}) when the acoustic cycle is neglected leads to:
\begin{eqnarray}
\omega_r(k)\tau_{\cal Q}+\phi_{\cal Q}=2(k+1)\pi.
\end{eqnarray}
A comparison of the frequency $\omega_r(k=0)$ of the fundamental mode with the frequency of the first harmonic $\omega_r(k=1)$, shown in Fig.~\ref{figphase}, indicates that $\omega_r(1)\sim 2\omega_r(0)$. We conclude that the phase of ${\cal Q}$ is negligible, and that the oscillation period of the fundamental mode is a good measure of the timescale $\tau_{\cal Q}$ at low frequency.\\

The fundamental mode is not always the most unstable one among the $l=1$ perturbations: Fig.~\ref{figa6b1wcut_wopt} summarizes the frequency range of unstable modes for $\alpha=3/2$, $\beta=5/2$ (upper plot) and $\alpha=6$, $\beta=1$ (lower plot). The most unstable mode may correspond to either the fundamental mode, the first or second harmonic as the shock radius increases from $r_\sh/r_\nabla=1.5$ to $100$. For both sets of cooling parameters, the corresponding oscillation period $2\pi/\omega_r$ of the most unstable $l=1$ mode would therefore be $\tau_{\cal Q}$ for $1.5<r_\sh/r_\nabla<5-6$, $\tau_{\cal Q}/2$ for $5-7<r_\sh/r_\nabla<33-39$, and even $\tau_{\cal Q}/3$ for $r_\sh/r_\nabla>33-39$. \\
An upper bound of the efficiency $|{\cal Q}|$ associated with the low frequency modes may be estimated from Eq.~(\ref{pureaa}) by neglecting the purely acoustic cycle and approximating $\tau_{\cal Q}\sim 2\pi (k+1)/\omega_r(k)$. The value of $\exp(\omega_i\tau_{\cal Q})$ is shown in Fig.~\ref{figQ0} for the first three eigenfrequencies ``f", ``h1", and ``h2", as a function of the shock radius. The actual value of ${\cal Q}$ at large shock radius is likely to be intermediate between the curve ``f" and the curve ``h1" in Fig.~\ref{figQ0} for $10\le r_{\sh}/r_\nabla\le 30$, as suggested by Fig.~\ref{figr5_30}: indeed, the eigenspectrum oscillation at low frequency in Fig.~\ref{figr5_30} suggests that the first harmonics ``h1" and ``h2" benefit from a constructive influence of the acoustic cycle, whereas this influence seems destructive on the fundamental mode ``f". In this respect, Fig.~\ref{figQ0} indicates that $|{\cal Q}|_{\rm wkb}$ can be used as  an acceptable guess of the efficiency $|{\cal Q}|$ at low frequency.\\
Besides, Fig.~\ref{figQ0} suggests a slow increase of the efficiency $|{\cal Q}|$ with the shock radius, with remarkable similarity for the two sets of cooling parameters (full and dashed lines).\\
 As the shock radius is increased, the slow increase of the oscillation frequency of the most unstable mode (Fig.~\ref{figa6b1wcut_wopt}), and the increase of $|{\cal Q}|$ (Fig.~\ref{figQ0}) are not explained yet.

\section{Continuity argument for the advective-acoustic instability at low frequency\label{sect_discuss}}

\begin{figure}
\plotone{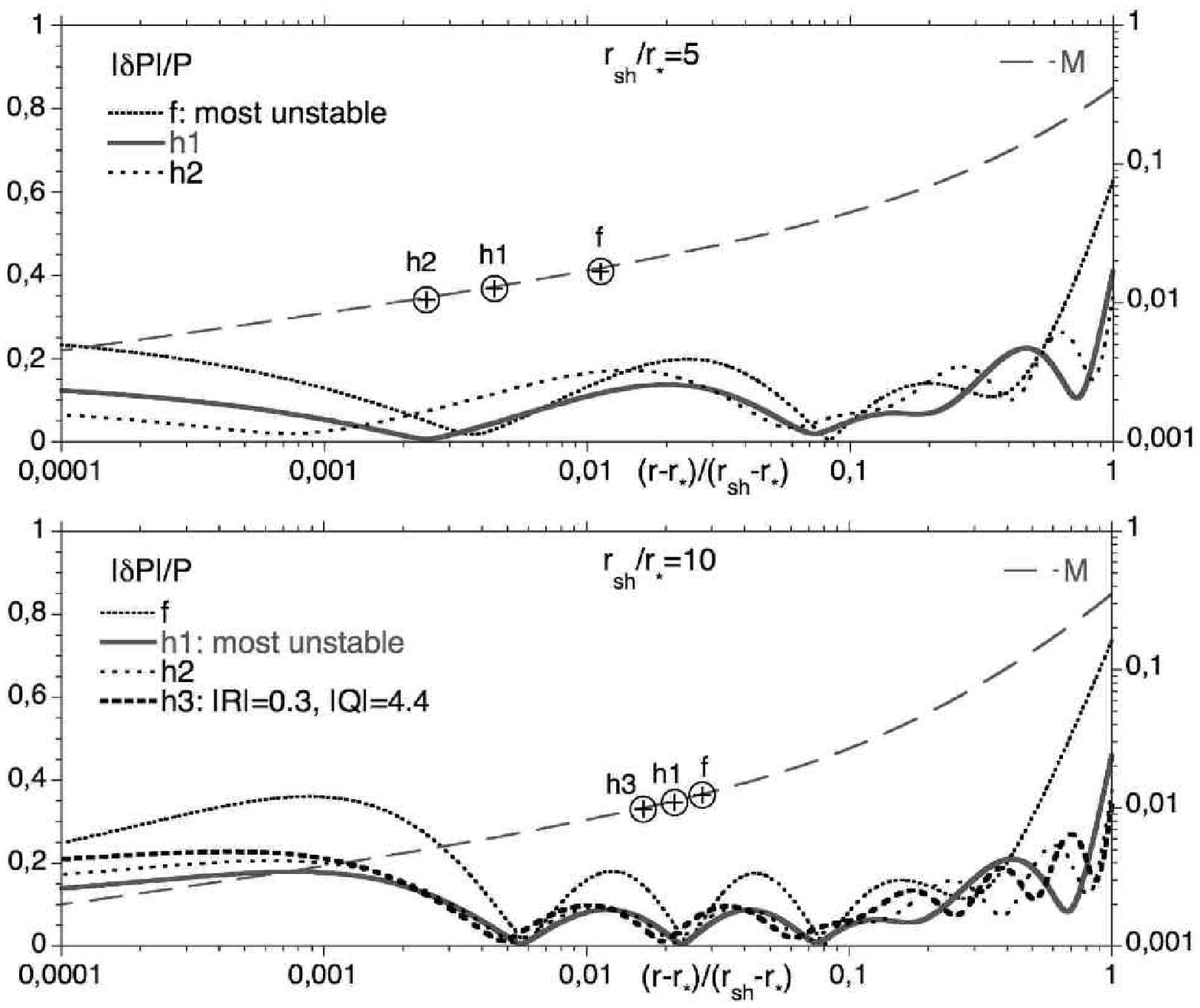}
\caption[]{Radial profiles of the pressure perturbation $|\delta P|/P$ for the fundamental $l=1$ mode (``f") and its first harmonics (``h1", ``h2"...), in the flow with $\alpha=3/2$, $\beta=5/2$, $r_\sh/r_*=5$ (upper plot) and $r_\sh/r_*=10$ (lower plot).
The pressure profile is shown as a function of the fractional radius, with the corresponding Mach number shown on the right axis. In the flow with $r_\sh/r_*=10$, the frequency of the third harmonic ``h3" is high enough to establish the instability of the advective-acoustic cycle $|{\cal Q}|=4.4$ and the stability of the purely acoustic cycle $|{\cal R}|=0.3$ (upper plot in Fig.~\ref{figa32b52QR}).
The shapes of all these eigenfunctions look very similar, suggesting a common physical mechanism. A lower bound of the position of the effective coupling radius $r_{\rm eff}$ is indicated by a circled symbol ``+", corresponding to $\tau_{\rm adv}(r)=2\pi(k+1)/\omega_r$ for the k$^{\rm th}$ harmonic. }
\label{figpression}
\end{figure}

Strictly speaking, we have demonstrated that the advective-acoustic cycle is responsible for the high frequency $l=1$ instability of a stalled accretion when it is far enough from the accretor, for two different types of cooling functions. What is the bearing of this demonstration on the instability mechanism of the low frequency modes, when the shock radius is moderate ? Since the analysis of the fundamental $l=1$ mode and its first harmonics is intractable through usual WKB techniques, one might argue that the instability mechanism of the fundamental mode is out of reach of the present study. The eigenspectrum of the flow with $\alpha=6$, $\beta=1$ and $r_\sh/r_\nabla=5$ (lower plot of Fig.~\ref{figr5_30}) contains too few unstable modes to allow for the identification of oscillations, and the frequency of these eigenmodes is too low to allow for a WKB analysis. In this case, neither of the two methods described in 
Sect.~\ref{sect_formalism} is able to compute $|{\cal Q}|$ and $|{\cal R}|$. 
Fig.~\ref{figr5_30}, however, shows the continuity of the shape of the eigenspectrum, both with respect to frequency and with respect to the shock radius. 
The growth rate of the low frequency eigenmodes is only marginally larger than those at higher frequency.  We feel there is no need to invoke a different instability mechanism at low frequency given the smooth distribution of growth rates. According to Fig.~\ref{figr5_30}, decreasing the shock radius seems to reduce the cut-off frequency above which the modes are stable, but barely affects the growth rate of the low frequency modes. The continuity of the flow properties with respect to the shock radius is also apparent in the sequence of plots in the Figs.~\ref{figa32b52QR}-\ref{figa6b1QR} and Figs.~\ref{figa32b52r9_29wi}-\ref{figa6b1r10_30wi}.

Besides, the radial structure of the pressure perturbation looks very similar for all the harmonics displayed in Fig.~\ref{figpression},  for two different shock radii. The eigenfunction of the low frequency, most unstable mode resembles those of higher frequency modes for which a reliable determination of the cycle efficiencies ${\cal Q}$ and ${\cal R}$ is possible. The similarity of the eigenfunctions suggests again a common instability mechanism for all these modes, namely the instability of the advective-acoustic cycle. 

Could the same kind of mechanism account for the slow instability of the fundamental $l=0$ mode, observed in Fig.~\ref{fig43a32b52wri} or \ref{fig43a32b52l06} for a large shock radius $r_\sh/r_*>15$ ?
The eigenspectrum associated to the radial mode shows oscillations very similar to Fig.~\ref{figr5_30}, except that all the higher harmonics are stable, even for a very large shock radius $r_\sh/r_*=1000$.
These oscillations enable the direct computation of the efficiencies $|{\cal Q}|$ and $|{\cal R}|$ of the acoustic cycle. The ``entropic-acoustic" cycle is always stable $|{\cal Q}|<1$ at the high frequencies required by the WKB analysis, 
and the purely acoustic cycle is also stable $|{\cal R}|<1$. Since the first harmonics seem always stable, 
the instability mechanism of the fundamental $l=0$ mode is more difficult to interpret than for $l=1$ modes. We cannot exclude, however, that it may be due to a low frequency entropic-acoustic cycle destabilized by the temperature increase in the adiabatic part of the flow, when the shock radius is large.

\section{Discussion of the acoustic interpretation of BM06\label{sect_BM06}}

BM06 summarized their preference for a purely acoustic mechanism by the following three observations:
\par (i) the advection time down to the accretor surface $\tau_{\rm adv}(r_*)$ is larger than the oscillation period $\tau_{\rm osc}\equiv2\pi/\omega_r$, whereas a non radial path exists such that the acoustic timescale along this path coincides with $\tau_{\rm osc}$ (Fig.~6 of BM06),
\par (ii) a standing $l=1$ acoustic wave is clearly visible on the pressure profile. The particular radius $r$ such that $\tau_{\rm adv}(r)\sim \tau_{\rm osc}$, which could have been identified as a radius of effective advective-acoustic coupling, does not show any particular signature on the pressure profile,
\par (iii) the instability seems independent of the flow near the accretor, given the resemblance between the results with the two different cooling functions which differ only in the most inner regions.\\
We believe that part of the confusion comes from the fact that for a post-shock Mach number $\M_\sh\sim 1/8^{1/2}$ typical of a strong adiabatic shock with $\gamma=4/3$, the advection timescale $r_\sh/v_\sh$ happens to be both longer than the radial acoustic timescale $2r_\sh/c_\sh$, and shorter than the surface acoustic time $2\pi r_\sh/c_\sh$:
\begin{eqnarray}
{2r_\sh\over c_\sh}<{r_\sh\over v_\sh}<{2\pi r_\sh\over c_\sh}.
\end{eqnarray}
An advective-acoustic timescale of the order of $r_\sh/v_\sh$ can thus be matched by an acoustic time along an ad-hoc nonradial path. This confusion between the advective and acoustic timescales could disappear if the postshock flow were very subsonic, with $\M_\sh\ll(2\pi)^{-1}=0.16$. Such conditions could be obtained in a flow involving a strong leak of energy at the shock, mimicking photodissociation in the same manner as Foglizzo, Scheck \& Janka (2006). Unfortunately, preliminary calculations indicate that the shock distance is so much diminished by this energy loss that the $l=1$ mode is stabilized.

We argue below that the observations of BM06 do not contradict our description of the advective-acoustic instability.
\par - The resemblance between the two flows with different cooling functions (point iii) has been investigated in depth throughout this paper, showing that the instability is dominated in both flows by the advective-acoustic cycle occurring between the shock and the effective coupling radius $r_{\rm eff}\sim r_\nabla$. Some significant differences were also discussed, such as the smaller frequency cut-off in the flow with $\alpha=6$, $\beta=1$ which could be due to the smoother profile of the velocity gradient (Fig.~\ref{figdvdr}).
\par - The existence of a standing pressure wave (point ii) cannot be held against the advective-acoustic cycle, as seen on Fig.~\ref{figpression}. Most of the radial structure of the perturbed pressure shown in Fig.~\ref{figpression} is very localized in the first $10\%$ of the shock distance, close to the accretor, and could be compatible with Fig.~7 of BM06 once projected on the $l=1$ spherical harmonics. Let us recall that $r_{\rm eff}$ is an ``effective" coupling radius, as defined in Sect.~\ref{sect_tauq}, and certainly not the unique radius of acoustic emission.   
\par - As discussed in Sect.~\ref{sect_tauq}, our estimate of $r_{\rm eff}\sim r_\nabla$ is not accurate enough to be a decisive test for our interpretation (point i). In a sense, our determination of $r_{\rm eff}$ in Fig.~\ref{figpression} could be considered as arbitrary as the determination of an ``acoustic" path in Fig.~6 of BM06. Nevertheless, independently of estimating $r_{\rm eff}$, we did prove the instability of the advective-acoustic cycle at high frequency in Sect.~\ref{sect_mechan}, and showed the similarity with low frequency modes in Sect.~\ref{sect_discuss}. 

In view of the above detailed description of the advective-acoustic instability, together with the proven stability of the purely acoustic cycle at high frequency, the possibility that a purely acoustic mechanism may be responsible for the most unstable low frequency modes seems unlikely. 

\section{Conclusions\label{sect_conc}}

This work is the first characterization, through a linear study, of the advective-acoustic instability in a decelerated accretion flow involving cooling processes. Our formulation of the boundary conditions at the shock corrects an error in HC92 concerning non radial perturbations. The numerical solution of this problem confirms the existence of an $l=1$ instability, as found in the numerical simulations of Blondin \etal (2003), for the two types of cooling functions studied by BM06. A detailed comparison of the growth time and the oscillation period of the dominant mode of the instability
revealed discrepancies which can reach $\sim30\%$, possibly due to the numerical difficulty of advecting vorticity waves towards the region of deceleration without artificially damping them by numerical viscosity. The optimal grid size near the accretor surface is estimated as a fraction $0.1-0.2\%$ of the shock distance from the accretor.  \\
The main purpose of this study was to clarify the instability mechanism at work for $l=1$ perturbations, with the following results for both types of considered cooling functions $\alpha=3/2$, $\beta=5/2$ and $\alpha=6$, $\beta=1$:
\par (1) We have proven, for the first time, that an advective-acoustic instability of the $l=1$ mode takes place in a decelerated accretion flow involving cooling processes. The WKB approximation used in this proof required that the shock radius exceeds $\sim 10$ times the accretor radius.

\par (2) The low frequency $l=1$ instability occurring when the shock distance is moderate ($r_\sh/r_\nabla\ge2$) has also been interpreted as an advective-acoustic instability, using our conclusion (1) together with a continuity argument: the instability of the low frequency modes can be interpreted in continuity with the instability at higher frequency in a series of flows with larger shock radii. This continuity argument is based on a comparison of both the eigenspectra (Fig.~\ref{figr5_30}) and the eigenfunctions (Fig.~\ref{figpression}) of the unstable modes.

\par (3) The purely acoustic cycle is very stable ($|{\cal R}|\le 0.5$) in the range of shock radii and frequencies allowed by our approximations. This result disfavours of the acoustic interpretation of BM06.

\par (4) The efficiency of the advective-acoustic cycle is an increasing function of the shock distance, which reaches an efficiency $|{\cal Q}|\sim 4-5$ for a shock radius $r_\sh/r_\nabla\sim 10$.

\par (5) We have proposed a simple approximation of the advective-acoustic cycle timescale $\tau_{\cal Q}$ based on $\tau_\nabla$, defined as the advection time from the shock to the radius $r_\nabla$ where the velocity gradient is strongest, plus a radial acoustic feedback for the sake of simplicity. This estimate is quite accurate at high frequency, but overestimates the cycle timescale at low frequency by $\sim 20\%$.

\par (6) We have shown that the oscillation period of the fundamental mode is a measure of the timescale $\tau_{\cal Q}$. The oscillation period of the most unstable mode is comparable to $\tau_{\cal Q}$,  $\tau_{\cal Q}/2$ or $\tau_{\cal Q}/3$ depending on the shock radius (Fig.~\ref{figa6b1wcut_wopt}). \\

Our efforts to understand the instability mechanism at work in a simplified flow aim at guiding our intuition when interpreting more complex numerical simulations of astrophysical flows (\ie Scheck \etal 2006b). The general description of the instability is globally satisfactory, but some fundamental questions are still unanswered. The following three questions may be answered by further studies:
\par (i) What is the maximum efficiency $|{\cal Q}|_{\rm max}$ of an advective-acoustic cycle with cooling ? We observed in Figs.~\ref{figQ0} that $|{\cal Q}|_{\rm max}$ increases slowly with the shock distance
 for both types of cooling function: is this due to the influence of an extended quasi adiabatic region where enthalpy gradients contribute to the advective-acoustic coupling (F01) ? Or is this related to the cooling mechanism in the deceleration region ? How does $|{\cal Q}|_{\rm max}$ depend on the adiabatic index $\gamma$ ? Answering these questions should help us estimate the efficiency $|{\cal Q}|$ in astrophysical flows with a more realistic equation of state and more elaborate cooling processes. 
\par (ii) What are the conditions for the dominance of a $l=2$ mode, as observed in Fig.~\ref{fig43a32b52l06} for $\alpha=3/2$, $\beta=5/2$ for $1.5<r_{\rm sh}/r_*<1.9$ ? How does the maximum efficiency $|{\cal Q}|_{\rm max}$ depend on the degree $l$ of the perturbation ? This question could be important with respect to the asymmetry of the explosion, and the subsequent kick of the neutron star (Scheck et al. 2006a).
\par (iii) Can we better understand the conditions under which the instability is dominated by the fundamental mode, its first or second harmonic ? This question may be important with respect to the explosion mechanism proposed by Burrows \etal (2006). This mechanism requires the nonlinear transfer of energy from an unstable flow above the neutron star to a gravity mode of the neutron star, whose frequency can be a factor 10 higher. We may expect that the higher the frequency of the advective-acoustic cycle, the easier the excitation of the gravity mode of the neutron star through nonlinear processes.\\
Some of these questions can be addressed by further simplifying the accretion flow in order to allow for analytical calculations of the advective-acoustic cycle, free from the low frequency limitations inherent to the WKB approximation. This is the purpose of a companion paper (Foglizzo \etal 2006).

\acknowledgments
The authors are grateful to the referee, R. Chevalier, for his constructive and helpful comments.
Funding by Egide (France) and by DAAD (Germany) through their ``Procope" exchange program, and
support by the Sonderforschungsbereich 375 on ``Astro-Particle Physics" of the Deutsche 
Forschungsgemeinschaft, are acknowledged. T.F. is thankful to the KITP for its stimulating program ``The supernovae-GRB connection", supported in part by the National Science Foundation under
Grant No. PHY99-07949

\appendix

\section{Explicit relations between $\delta v$, $\delta\rho$,  $\delta P$, $\delta {\cal L}$ and $f$, $h$, $\delta K$}

The functions $f$, $h$, $\delta S$ can be translated into the classical variables $\delta v_r$, $\delta \rho$, $\delta P$, $\delta c^2$ using Eqs.~(\ref{defS}), (\ref{deff}) and (\ref{defg}):
\begin{eqnarray}
{\delta v_r\over v}&=&{1\over 1-\M^2}\left(h+\delta S-{f\over c^2}\right),\label{dvv}\\
{\delta \rho\over\rho}&=&{1\over 1-\M^2}\left(-\M^2h-\delta S+{f\over c^2}\right),\label{drho}\\
{\delta P\over \gamma P}&=&{1\over 1-\M^2}\left\lbrack
-\M^2h-(1+(\gamma-1)\M^2){\delta S\over\gamma}
+{f\over c^2}\right\rbrack,\label{dpp}\\
{\delta c^2\over c^2}&=&{\gamma-1\over 1-\M^2}\left({f\over c^2}-\M^2{ h}
-\M^2{\delta S}\right).\label{dcc}
\end{eqnarray}
The functions $\delta K$, $f$ are related to the transverse velocity perturbation $(0,\delta v_\theta,\delta v_\varphi)$, according to the following equation, obtained from a combination of the transverse components of the Euler equation:
\begin{eqnarray}
\delta A&\equiv&{r\over\sin\theta}\left\lbrack{\p\over\p\theta}(\sin\theta\delta v_\theta)+{\p\over\p\varphi}\delta v_\varphi\right\rbrack\label{defA},\\
&=&{1\over i\omega}\left\lbrack\delta K-l(l+1)f\right\rbrack.\label{AK}
\end{eqnarray}
Using the fact that $P=\rho c^2/\gamma$, the heating function defined by Eq.~(\ref{defL}) and Eq.~(\ref{eqS}) is perturbed as follows:
\begin{eqnarray}
\delta\left({{\cal L}\over \rho v}\right)&=&{\nabla S}{c^2\over \gamma}\left\lbrack
(\beta-1){\delta\rho\over\rho}+\alpha{\delta c^2\over c^2}-{\delta v_z\over v}
\right\rbrack,\\
\delta\left({{\cal L}\over pv}\right)&=&{\gamma\over c^2}\delta\left({{\cal L}\over \rho v}\right)
-{\delta c^2\over c^2}{\nabla S},
\end{eqnarray}
In these equations, the perturbations $\delta v_z$, $\delta\rho$ and $\delta c^2$ can be replaced by functions of $f,h,\delta S$ using Eqs.~(\ref{dvv}-\ref{dcc}).

\section{Shock boundary conditions}

The boundary condition at the shock follows the conservation of mass flux, 
momentum flux and energy flux in the frame of the shock:
\begin{eqnarray}
\rho_{1}(v_{1}-\Delta v)&=&(\rho_{\sh}+\delta \rho_\sh)(v_{\sh}+\delta v_{\sh}-\Delta v),\\
\rho_{1}(v_{1}-\Delta v)^{2}+\rho_{1}{c_{1}^{2}\over\gamma}&=&
(\rho_{\sh}+\delta \rho_\sh)(v_{\sh}+\delta v_{\sh}-\Delta v)^{2}\nonumber\\
&&+(\rho_{\sh}+\delta \rho_\sh){(c_{\sh}+\delta c_{\sh})^{2}\over\gamma},\\
{(v_{1}-\Delta v)^{2}\over2}+{c_{1}^{2}\over\gamma-1}&=&
{(v_{\sh}+\delta v_{\sh}-\Delta v)^{2}\over2}
+{(c_{\sh}+\delta c_{\sh})^{2}\over\gamma-1},
\end{eqnarray}
where quantities are measured at the position $\rsh+\Delta\zeta$. Keeping 
the first order terms, and using the definition of $f,h$, these equations are rewritten at the 
position $\rsh$ using a Taylor expansion:
\begin{eqnarray}
\rho_{1}v_{1}h_\sh-(\rho_{\sh}-\rho_{1})\Delta v=
\Delta\zeta\left\lbrack
{\p\over\p r}\left(\rho v\right)_1
-{\p\over\p r}\left(\rho v\right)_\sh
\right\rbrack,\\
v_{\sh}^2\delta \rho_\sh+2\rho_\sh v_\sh\delta v_\sh
+{2\over\gamma}\rho_{\sh}c_{\sh}\delta c_{\sh}+\delta \rho_\sh {c_{\sh}^2\over\gamma}
=\nonumber\\
\Delta\zeta\left\lbrack
{\p\over\p r}\left(\rho v^2+P\right)_1
-{\p\over\p r}\left(\rho v^2+P\right)_\sh
\right\rbrack
,\\
f_\sh-(v_\sh-v_{1})\Delta v=\nonumber\\
\Delta\zeta\left\lbrack
{\p\over\p r}\left({v^2\over2}+{c^2\over\gamma-1}\right)_1
-{\p\over\p r}\left({v^2\over2}+{c^2\over\gamma-1}\right)_\sh
\right\rbrack.
\end{eqnarray}
From the equations (\ref{eqcont}), (\ref{eqbern}), (\ref{eqS}) of the stationary flow,
\begin{eqnarray}
{\p\over\p r}\left(\rho v\right)&=&-2{\rho v\over r},\\
{\p\over\p r}\left(P+\rho v^2\right)&=&-\rho{GM\over r^2}-2{\rho v^2\over r},\label{gradpress}\\
{\p\over\p r}\left({v^2\over2}+{c^2\over\gamma-1}\right)&=&{{\cal L}\over\rho v}-{GM\over r^2}.
\end{eqnarray}
We obtain:
\begin{eqnarray}
h_\sh&=&\left({1\over v_{\sh}}-{1\over v_{1}}\right)\Delta v,\label{boundg}\\
{\delta S_\sh\over\gamma}&=&
{\Delta\zeta\over c_\sh^2} \left\lbrack {{\cal L}_1\over\rho_1 v_1}
-{{\cal L}_\sh\over\rho_\sh v_\sh}-\left({GM\over r_\sh^2}-2{v_1v_\sh\over r_\sh}\right)\left(1-{v_\sh\over v_1}\right)\right\rbrack\nonumber\\
&&-{v_1\Delta v\over c_\sh^2}\left(1-{v_\sh\over v_1}\right)^2,\label{dssh}\\
f_\sh&=&(v_{\sh}-v_{1})\Delta v+\Delta\zeta\left({{\cal L}_1\over\rho_1 v_1}
-{{\cal L}_\sh\over\rho_\sh v_\sh}\label{fshapp}
\right).
\end{eqnarray}
In the entropy equation, the gravity term  $GM/r_\sh^2$ and the term $v_1v_\sh/r_\sh$ due to the spherical geometry can be rewritten as follows:
\begin{eqnarray}
{GM\over r_\sh^2}-2{v_1v_\sh\over r_\sh}&=&{v_{\rm ff}^2\over 2r_\sh}-2{v_1v_\sh\over r_\sh},\\
&=&{v_1^2\over 2r_\sh}\left\lbrack\left({v_{\rm ff}\over v_1}\right)^2-4{v_\sh\over v_1}\right\rbrack.
\end{eqnarray}
The transverse velocity immediately after the shock is deduced from the conservation of the tangential 
component of the velocity, in the spirit of Landau \& Lifschitz (1989), leading to Eqs.~(\ref{dvtheta}-\ref{dvphi}). $\delta A_\sh$ is deduced from its definition (\ref{defA}) and Eqs.~(\ref{dvtheta}-\ref{dvphi}):
\begin{eqnarray}
\delta A_{\rm sh}&=&-l(l+1)(v_1-v_{\rm sh})\Delta\zeta.
\end{eqnarray}
$\delta K_{\sh}$ is deduced from $\delta A_\sh$ using Eq.~(\ref{AK}) and (\ref{fshapp}).
The assumption that ${\cal L}_1\ll {\cal L}_\sh$, with ${\cal L}_\sh=\rho_\sh v_\sh c_\sh^2\nabla S_\sh/\gamma$, leads to Eqs.~(\ref{fsh2}), (\ref{hsh2}) and (\ref{Ssh2}) if the shock is strong.

\section{Projection of perturbations on acoustic and advected waves}

\subsection{Uniform adiabatic flow}

In a uniform, adiabatic flow moving at constant velocity in the direction $z$, any perturbation $f,h,\delta S,\delta K$ associated with the frequency $\omega$ and perpendicular wavenumber $k_\perp$ can be decomposed as a sum of acoustic waves and advected waves as follows:
\begin{eqnarray}
f&=&f^++f^-+f^S+f^K,\\
h&=&h^++h^-+h^S+h^K,
\end{eqnarray}
where an acoustic wave is noted $f^+,h^+$ if it propagates in the direction of the flow, and $f^-,h^-$ otherwise. The contribution to $f,h$ of a vorticity perturbation $\delta K$ such that $\delta S=0$ is:
\begin{eqnarray}
f^K&=&{\M^2(1-\mu^2)\over 1-\mu^2\M^2}{\delta K\over k_\perp^2},\label{deffK}\\
h^K&=&{f^K\over v^2},\label{defhK}
\end{eqnarray}
where $\mu^2\equiv 1-k_\perp^2c^2(1-\M^2)/\omega^2$.
An entropy-vorticity perturbation such that $\delta K=0$ contributes to the perturbation $f,h$ as follows:
\begin{eqnarray}
{f^S\over c^2}&=&{1-\M^2\over 1-\mu^2\M^2}{\delta S\over \gamma}\label{deffS},\\
h^S&=&{\mu^2\over c^2}f^S-\delta S.\label{defhS}
\end{eqnarray}
The acoustic component in a uniform adiabatic flow is deduced from $f,h,\delta S,\delta K$ through:
\begin{eqnarray}
f^{\pm}&=& {1\over2}f \pm {{\cal M}c^2\over2\mu}(h+\delta S)-
{1\pm\mu\M\over2}\left(f^S
\pm{f_K\over\mu\M}\right) ,\label{deffpm}\\
h^{\pm}&=&\pm {\mu\over\M}{f^{\pm}\over c^2}.\label{defhpm}
\end{eqnarray}

\subsection{Extension to a spherical flow with cooling}

In a spherical flow with gradients, the advected and propagating waves are no longer independent, but coupled even if the flow were adiabatic. Moreover, cooling processes are responsible for an additional coupling between advected and propagating perturbations. 
We choose to use the same decomposition obtained in a uniform adiabatic flow, adapted to spherical coordinates by replacing $\mu^2$ by the spherical value (Eq.~\ref{defmu}), and $k_\perp^2$ by $l(l+1)/r^2$ (the eigenvalue of the Laplacian operator):
\begin{eqnarray}
f^K&\equiv&{\M^2(1-\mu^2)\over 1-\mu^2\M^2}{\delta K\over l(l+1)},\label{deffKsp}\\
h^K&\equiv&{f^K\over v^2},\label{defhKsp}\\
{f^S\over c^2}&\equiv&{1-\M^2\over 1-\mu^2\M^2}{\delta S\over \gamma}\label{deffSsp},\\
h^S&\equiv&{\mu^2\over c^2}f^S-\delta S.\label{defhSsp}\\
f^{\pm}&\equiv& {1\over2}f \pm {{\cal M}c^2\over2\mu}(h+\delta S)-
{1\pm\mu\M\over2}\left(f^S
\pm{f_K\over\mu\M}\right) ,\label{deffpmsp}\\
h^{\pm}&\equiv&\pm {\mu\over\M}{f^{\pm}\over c^2}.\label{defhpmsp}
\end{eqnarray}
For any perturbation $f,h,\delta S,\delta K$ in a spherical nonadiabatic flow, the quantities ($f^\pm,f^K,f^S$) and ($h^\pm,h^K,h^S$) defined above naturally satisfy:
\begin{eqnarray}
f&=&f^++f^-+f^S+f^K,\\
h&=&h^++h^-+h^S+h^K.
\end{eqnarray}
This decomposition describes the amount of advected and propagating waves that would be measured if the perturbation were allowed to continue its evolution in a uniform adiabatic flow. We choose to apply this decomposition at the shock radius, on the subsonic side. The identification of $f^\pm$ with acoustic waves is strictly valid in the WKB approximation, when the wavelength is short compared to the scale of the gradients in the flow. The threshold of validity of this approximation is evaluated in Sect.~\ref{sect_QRWKB}.

\section{Numerical procedure to calculate ${\cal R}_\nabla$, ${\cal Q}_\nabla$ and ${\cal Q}^K_\nabla$}

One consequence of cooling is the fact that vorticity and entropy perturbations produced at the shock do not satisfy $\delta K_{\rm sh}\equiv0$ as in adiabatic flows (Eq.~\ref{Kshsp}). This raises the question of the acoustic feedback produced by the advection of the residual value of $\delta K_\sh$ generated by cooling at the shock. The global efficiency ${\cal Q}^K$ associated with this feedback is thus also computed as a check of consistency: this feedback is indeed small in all our calculations 
($|{\cal Q}^K|<0.1$).

\subsection{Numerical calculation of ${\cal R}_\nabla$, ${\cal Q}_\nabla$, ${\cal Q}^K_\nabla$}

For a given frequency $\omega_r$ and degree $l$, the differential system is integrated four times from $r_{\rm sh}$ to $r_*$ with the following boundary conditions:
\par (i) acoustic wave propagating downward:
\begin{eqnarray}
\delta K_\sh=0,\; \delta S_\sh=0,\;f_\sh=f^+_\sh=1,\;h_\sh={\mu_\sh\over\M_\sh}{f_\sh\over c_\sh^2}.
\end{eqnarray}
\par (ii) acoustic wave propagating upward:
\begin{eqnarray}
\delta K_\sh=0, \;\delta S_\sh=0,\;f_\sh=f^-_\sh=1,\;h_\sh=-{\mu_\sh\over\M_\sh}{f_\sh\over c^2}.
\end{eqnarray}
\par (iii) entropy/vorticity wave advected downward:
\begin{eqnarray}
\delta K_\sh=0, \;\delta S_\sh=1,\;f_\sh=f_\sh^S,\;h_\sh=h_\sh^S.
\end{eqnarray}
\par (iv) vorticity wave advected downward:
\begin{eqnarray}
\delta K_\sh=1, \;\delta S_\sh=0,\;f_\sh=f_\sh^K,\;h_\sh=h_\sh^K.
\end{eqnarray}
In each of these four cases, the differential system is integrated from the shock down to the accretor surface $r_*$, where the velocity perturbation reaches a value $(\delta v/v)(r_*)$ noted $a_+$, $a_-$, $a_S$ and $a_K$ respectively. A linear combination of a couple of these integrated solutions allows us to construct three solutions which fulfills the boundary condition $(\delta v/v)(r_*)=0$ and measure at the shock radius the following efficiencies of acoustic feedback within the flow:
\par (i) Acoustic reflection, without any advected perturbation at the shock:
\begin{eqnarray}
{\cal R}_\nabla\equiv-{a_+\over a_-}.
\end{eqnarray}
\par (ii) Acoustic feedback produced by an entropy/vorticity perturbation such that $\delta K_{\rm sh}=0$:
\begin{eqnarray}
{\cal Q}_\nabla\equiv-{1\over a_-}{a_S\over f^S}.
\end{eqnarray}
\par (iii) Acoustic feedback produced by a vorticity perturbation such that $\delta S_{\rm sh}=0$:
\begin{eqnarray}
{\cal Q}_\nabla^K\equiv-{1\over a_-}{a_K\over f^K}.
\end{eqnarray}

\subsection{Calculation of ${\cal R}_\sh$, ${\cal Q}_\sh$, ${\cal Q}^K_\sh$}

At the shock, the coupling coefficients ${\cal R}_{\rm sh}$, ${\cal Q}_{\rm sh}$ and ${\cal Q}_{\rm sh}^K$ are defined by
\begin{eqnarray}
{\cal R}_{\rm sh}&\equiv&{f^+_{\rm sh}\over f^-_{\rm sh}},\\
{\cal Q}_{\rm sh}&\equiv&{f^S_{\rm sh}\over f^-_{\rm sh}},\\
{\cal Q}_{\rm sh}^K&\equiv&{f^K_{\rm sh}\over f^-_{\rm sh}},
\end{eqnarray}
where $f^\pm_{\rm sh}$, $f^S_{\rm sh}$ and $f^K_{\rm sh}$ are deduced from Eqs.~(\ref{deffK}), (\ref{deffS}) and (\ref{deffpm}), with the boundary values $f_{\rm sh}$, $h_{\rm sh}$, $\delta S_{\rm sh}$ and $\delta K_{\rm sh}$ established in Eqs.~(\ref{fsh2}-\ref{Kshsp}). Although cooling is neglected in the projection of the perturbation $f$ on $f^\pm,f^K,f^S$ (Eqs.~\ref{deffKsp}-\ref{defhpmsp}), some effect of cooling on the jump conditions is taken into account through $f^\pm_{\rm sh}$, $f^S_{\rm sh}$ and $f^K_{\rm sh}$.


\begin{thebibliography}{}

\bibitem[]{akt84}
Abouseif, G.E., Keklak, J.A., Toong, T.Y., 1984, Combustion Science and 
Technology, 36, 83

\bibitem[]{b86}
Bertschinger, E. 1986, \apj, 304, 154

\bibitem[2003]{bmd03}
Blondin, J.M, Mezzacappa, A., \& DeMarino, C. 2003,  \apj, 584, 971

\bibitem[2003]{bm06}
Blondin, J.M, \& Mezzacappa, A. 2006,  \apj, 642, 401 (BM06)

\bibitem[]{b05}
Burrows, A., Livne, E., Dessart, L., Ott, C.D., \& Murphy, J., 2006, \apj, 640, 878

\bibitem[]{c72}
Candel, S.M., 1972, PhD Thesis, California Institute of Technology, 
Pasadena, California. ``Analytical studies of some acoustic problems of 
jet engines".

\bibitem[]{ci82}
Chevalier, R.A., \& Imamura, J.N. 1982, \apj, 261, 543

\bibitem[2001]{f01}
Foglizzo, T. 2001, \aap, 368, 311 (F01)

\bibitem[2002]{f02}
Foglizzo, T. 2002, \aap, 392, 353 (F02)

\bibitem[2000]{ft00}
Foglizzo, T., \& Tagger, M. 2000, \aap, 363, 174

\bibitem[2003]{fg03}
Foglizzo, T., \& Galletti, P. 2003, proceeding of the conference ``Cosmic explosions in three dimensions", edited by P. Hoflich, P. Kumar and J.C. Wheeler, CUP, astro-ph/0308534

\bibitem[2006]{fsj06}
Foglizzo, T., Galletti, P., \& Ruffert, M. 2005, \aap, 435, 397

\bibitem[2006]{fsj06}
Foglizzo, T., Scheck, L., \& Janka, H.T. 2006, \apj, 652, in press, astro-ph/0507636

\bibitem[1992]{hc92}
Houck, J.C., \& Chevalier, R.A. 1992, \apj, 395, 592 (HC92)

\bibitem[]{h75}
Howe, M.S., 1975, J. Fluid Mech., 71, 625

\bibitem[]{iaww96}
Imamura, J.N., Aboasha, A., Wolff, M.T., Wood, K.S. 1996, \apj, 458, 327

\bibitem[]{jskmp04}
Janka, H.T., Scheck, L., Kifonidis, K., M\"uller, E., Plewa, T. 2004, proceedings of the conference ``The fate of the most massive stars", Humphreys \& Stanek (eds), ASP, p. 372, astro-ph/0408439

\bibitem[]{ll89} 
Landau, L., \& Lifschitz, E. 1989, Fluid Mechanics, Editions MIR

\bibitem[]{mc77}
Marble, F.E., \& Candel, S.M., 1977, Journal of Sound and Vibration, 55, 225

\bibitem[]{oky06}
Ohnishi, N., Kotake, K., \& Yamada, S. 2006, \apj, 641, 1018

\bibitem[]{s02}
Saxton, C.J. 2002, \pasa, 19, 282

\bibitem[]{sw99}
Saxton, C.J., \& Wu, K. 1999, \mnras, 310, 677

\bibitem[]{sfkj06}
Scheck, L., Foglizzo, T., Janka, H.T., Kifonidis, K.  2006b, \aap, to be submitted

\bibitem[]{skjm06}
Scheck, L., Kifonidis, K., Janka, H.T., M\"uller, E. 2006a, \aap, in press, astro-ph/0601302

\bibitem[]{spjm04} 
Scheck, L., Plewa, T., Janka, H.-T., \& M\"uller, E. 2004, \prl, 92, 1

\end{thebibliography}
\end{document}